\documentclass[aps,revtex,prm,twocolumn,amsmath,superscriptaddress,showpacs,floatfix,reprint]{revtex4-2}
\usepackage{amsmath, nccmath}
\usepackage{amssymb}
\usepackage{bm}
\usepackage{braket}
\usepackage{natbib}
\usepackage{leftidx}
\usepackage{graphicx}
\usepackage{verbatim}
\usepackage{subfigure}
\usepackage{hyperref}
\usepackage{dcolumn}
\usepackage{textcomp}
\usepackage{float}
\usepackage{threeparttable}
\usepackage{titlecaps}
\usepackage{multirow}
\usepackage{booktabs}
\usepackage{tabularx}
\usepackage{array}
\usepackage{lipsum}
\hypersetup{
	colorlinks=true,
	citecolor=blue,
	filecolor=black,
	linkcolor=blue,
	urlcolor=cyan
}
\usepackage{mathastext}
\usepackage{mathptmx}
\usepackage{enumitem}
\DeclareGraphicsExtensions{.png,.pdf,.tif}
\usepackage{pict2e}
\usepackage[dvipsnames]{xcolor}
\usepackage[normalem]{ulem}
\usepackage{comment}
\usepackage{multirow}
\usepackage{booktabs}
\usepackage{amsmath}

\begin{document}

\title{Tunable Magnetic Frustration in the Cu-Ru-based Double Perovskite \\La$_{2-x}$Sm$_x$CuRuO$_6$ (x = 0, 1, 2) Oxides}

\author{Soumya Ghorai}
\affiliation{Department of Condensed Matter and Materials Physics, S. N. Bose National Centre for Basic Sciences, Kolkata, West Bengal-700106, India.}

\author{Samir Rom}
\affiliation{Department of Condensed Matter and Materials Physics, S. N. Bose National Centre for Basic Sciences, Kolkata, West Bengal-700106, India.}

\author{Irina Shamova}
\affiliation{M. V. Lomonosov Moscow State University, Moscow 119991, Russia.}
\affiliation{National University of Science and Technology MISiS, Moscow 119049, Russia.}

\author{\textcolor{black}{N. K. Karn}}
\affiliation{\textcolor{black}{CSIR-National Physical Laboratory, Dr. K. S. Krishnan Marg, New Delhi, 110012, India.}}
\affiliation{\textcolor{black}{Academy of Scientific and Innovative Research (AcSIR), Ghaziabad, 201002, India.}}

\author{\textcolor{black}{Tamanna Kumari}}
\affiliation{\textcolor{black}{CSIR-National Physical Laboratory, Dr. K. S. Krishnan Marg, New Delhi, 110012, India.}}

\author{\textcolor{black}{A. K. Shukla}}
\affiliation{\textcolor{black}{CSIR-National Physical Laboratory, Dr. K. S. Krishnan Marg, New Delhi, 110012, India.}}
\affiliation{\textcolor{black}{Academy of Scientific and Innovative Research (AcSIR), Ghaziabad, 201002, India.}}

\author{Sanjoy Kr Mahatha}
\affiliation{UGC-DAE Consortium for Scientific Research, University Campus, Khandwa Road, Indore-452001, India.}

\author{O. Volkova}
\affiliation{M. V. Lomonosov Moscow State University, Moscow 119991, Russia.}

\author{Nitesh Kumar}
\affiliation{Department of Condensed Matter and Materials Physics, S. N. Bose National Centre for Basic Sciences, Kolkata, West Bengal-700106, India.}

\author{Tanusri Saha Dasgupta}
\email{t.sahadasgupta@gmail.com}
\affiliation{Department of Condensed Matter and Materials Physics, S. N. Bose National Centre for Basic Sciences, Kolkata, West Bengal-700106, India.}

\author{Setti Thirupathaiah}
\email{setti@bose.res.in}
\affiliation{Department of Condensed Matter and Materials Physics, S. N. Bose National Centre for Basic Sciences, Kolkata, West Bengal-700106, India.}

\date{\today}

\begin{abstract}
In this study, we investigate structural, magnetic, and electronic properties of the copper-ruthenate based oxide double perovskite La$_{2-x}$Sm$_x$CuRuO$_6$ (x = 0, 1, 2), synthesized through the solid-state reaction method. X-ray diffraction analysis reveals that all compounds crystallize in a monoclinic symmetry, with varying degree of structural distortion that increases in moving from La$^{3+}$ to smaller size cation Sm$^{3+}$. Electrical resistivity studies indicate insulating behaviour in all compounds, with variable-range-hopping domination at low temperatures, due to presence of anti-site disorder. AC susceptibility and heat capacity measurements suggest suppression of frustration in Sm-bearing compounds, affecting the magnetic behavior. Our first-principles calculations suggest that the combined effects of lattice distortion and Sm magnetism play a crucial role in weakening magnetic frustration, thereby rationalizing the experimental observations. These findings shed light on the complex interplay of crystal structure and magnetism in Cu-Ru double perovskites, and open up an avenue for tuning of magnetic properties through rare-earth-ion substitution.
\end{abstract}

\maketitle

\section{Introduction}\label{1}
\textcolor{black}{Magnetic frustration, where competing spin interactions hinder a system from achieving a unique ground state, is central to realizing exotic quantum phases~\cite{Lacroix2011}. The suppression of conventional long-range order in such systems often leaves fingerprints in low-temperature dynamical properties~\cite{Harris1997}. Geometrically frustrated lattices such as two-dimensional triangular and kagome arrangements~\cite{Han2012,Mourigal2014}, or three-dimensional pyrochlore and hyperkagome networks~\cite{Okamoto2007,Ross2011} have been instrumental in uncovering phenomena like spin liquids and spin glasses.}

\textcolor{black}{In this broader context, the perovskite-derived $A_2BB'O_6$ double perovskites provide a tunable platform for realizing complex magnetic phases driven by the interplay of lattice, charge, spin, and orbital degrees of freedom~\cite{Anderson1993,SahaDasgupta2020,Dass2004,Vasala2015}. The presence of transition metal ions at the $B$ and $B'$ sites leads to diverse phenomena, including colossal magnetoresistance~\cite{Kobayashi1998,Sarma2000prl}, multiferroicity~\cite{Rogado2005,Das2008}, and high spin polarization~\cite{Leng2022}. These compounds typically adopt a rock-salt type $B/B'$ ordering, forming two interpenetrating face-centered cubic (fcc) sublattices. When magnetic ions occupy these sites, the resulting topology of edge-sharing tetrahedra (as illustrated in Fig. 1 of the supplementary materials (SM)~\cite{SM}) inherently induces magnetic frustration under antiferromagnetic (AFM) interactions~\cite{Greedan2010}.}

\textcolor{black}{Within this family, we focus on copper-ruthenium (Cu-Ru) based double perovskites, where significant differences in the ionic size and charge of the localized Cu$^{2+}$ ($3d^9$) and extended Ru$^{4+}$ ($4d^4$) ions favour distinct site occupation~\cite{Greedan2010}. The parent compound, La$_2$CuRuO$_6$, is reported as ferrimagnet that notably exhibits spin-glass-like characteristics~\cite{Kumar2012,Panda2016}. Because the differing spatial extents of the $3d$ and $4d$ orbitals make these systems highly sensitive to chemical substitution, numerous $B$-site variants have been explored~\cite{Manna2016, Zhang2019,Dass2004}. Concurrently, $A$-site rare-earth substitutions (e.g., Pr$_2$CuRuO$_6$~\cite{Wang2021}, Nd$_2$CuRuO$_6$~\cite{Chen2021}, and Sm$_2$CuRuO$_6$~\cite{Chen2021}) have been shown to preserve the monoclinic (P2$_1$/n) structure while exhibiting ferromagnetic transitions. However, these prior studies primarily rely on basic temperature-dependent magnetization measurements and do not address the fate of the spin-glass dynamics upon rare-earth substitution.}

In order to bridge this gap, we synthesized La$_2$CuRuO$_6$ and two Sm substituted compounds, LaSmCuRuO$_6$ and Sm$_2$CuRuO$_6$, to systematically study the progressive influence of rare-earth element substitution \textcolor{black}{on the magnetic properties of La$_2$CuRuO$_6$, in particular,} the emergence or suppression of spin-glass-like dynamics. For this purpose, we employed a combination of experimental techniques, including X-ray diffraction (XRD), X-ray photoelectron spectroscopy (XPS), DC \& AC magnetization measurements, heat capacity measurement, and resistivity measurement. DC magnetization measurements show the enhancement of magnetic interaction by substitution of La with Sm. Whereas, the AC susceptibility measurements show the presence and absence of frequency dependence in La$_2$CuRuO$_6$ and Sm$_2$CuRuO$_6$, respectively. Heat capacity measurements further corroborate that the short range interaction present in La$_2$CuRuO$_6$ is suppressed in Sm$_2$CuRuO$_6$. The first-principles calculations supplement the experimental study by providing the microscopic origin of this trend. Our study reveals that the second nearest neighbour direct Ru-Ru interactions between extended 4$d$ states of Ru are large, and of antiferromagnetic (AFM) nature. The frustration arising from this AFM interaction in nearly equal length Ru-Ru bonds over the face of the tetrahedral unit of the double perovskite structure is responsible for the experimentally observed glassy behaviour in La$_{2}$CuRuO$_6$. On the other hand, replacement of La by Sm, makes the structure distorted with Ru-Ru bonds significantly unequal, thereby weakening the frustration drastically, as evident from the calculated Ru-Ru magnetic exchanges using the first-principles DFT. The dominant role of Ru-Ru interaction has been verified by considering hypothetical compounds, replacing the magnetic Cu$^{2+}$ ion by the non-magnetic Zn$^{2+}$ ion. It is to be noted that Sm, as opposed to La, is found to process a finite magnetic moment of $\sim$ 2.72-2.10 $\mu_B$, thus opening up \textcolor{black}{additional magnetic exchange pathways involving Sm, namely Cu-Sm and Ru-Sm exchange interactions. The inclusion of these additional
contribution in a highly sensitive magnetically frustrated system, becomes effective in relieving
the frustration effect. This together with lattice
distortion effect makes Sm$_2$CuRuO$_6$ significantly less frustrated compared to La$_2$CuRuO$_6$, thereby weakening the experimentally observed spin-glass signatures.}
Our findings highlight the tunability of magnetic properties in insulating double perovskites through A cation substitution.

\section{Methodology}\label{2}

\subsection{Experimental Details}
Polycrystalline powders of La$_{2}$CuRuO$_{6}$ (LCRO), LaSmCuRuO$_{6}$ (LSCRO), and Sm$_{2}$CuRuO$_{6}$ (SCRO) were synthesized using the conventional solid-state reaction method. Stoichiometric amount of high-purity ($\geq$ 99.9\%) reagents Sm$_{2}$O$_{3}$, La$_{2}$O$_{3}$, CuO, RuO$_{2}$ were used for the synthesis. These reagents were mixed using an agate mortar and pestle for almost 10 hours. The mixture was then placed in an alumina crucible and initially heated at 600$^\circ$C for 24 hours to minimize Ru evaporation. This was followed by a high-temperature heat treatment at 1100$^\circ$C for LCRO, and at 1000$^\circ$C for both LSCRO and SCRO, each for 24 hours. The samples were sufficiently ground for several hours after each heat treatment. We continued this process until the phase purity was achieved.

\begin{figure*}[]
	\includegraphics[width=0.75\linewidth, clip=true]{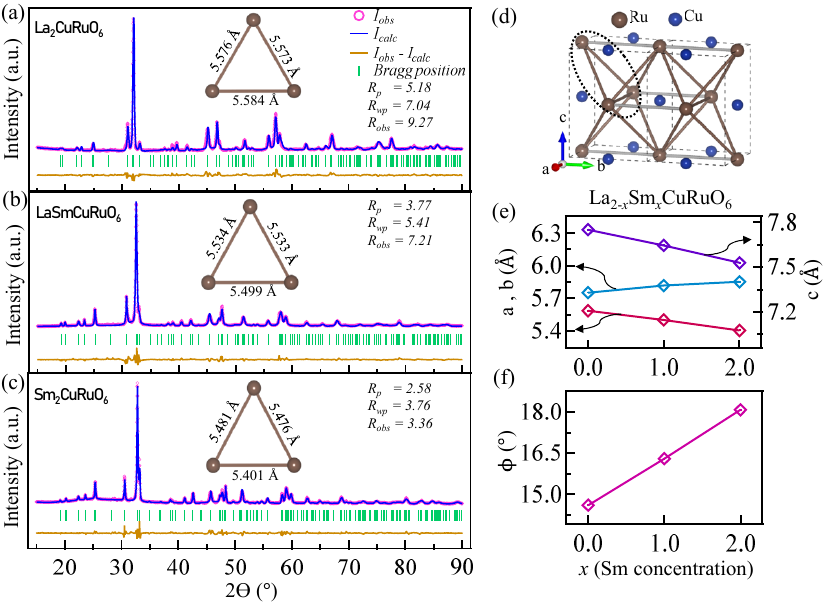}
	\caption{Rietveld refinement profile of the powder X-ray diffraction pattern for (a) La$_{2}$CuRuO$_{6}$ (b) LaSmCuRuO$_{6}$ (c) Sm$_{2}$CuRuO$_{6}$. Figure (d) shows the edge-sharing tetrahedral network of Ru sublattice. \textcolor{black}{Triangle formed by the second nearest neighbouring Ru ions are shown in the inset of each panel of (a)-(c). Figure (e) and (f) show the change of lattice parameters (a, b, c) and tilt angle ($\phi$) with Sm concentration.}}
	\label{fig2}
\end{figure*}

The phase purity of the as-grown powder samples was structurally analyzed using the high-resolution X-ray diffractometer (Rigaku smartLab 9 kW) having Cu-K$_\alpha$ radiation. The obtained powder X-ray diffraction patterns were analysed via the Rietveld refinement using \textsc{Jana-2020} software~\cite{Petricek2023} and three-dimensional visualization of the refined crystal structures generated using VESTA~\cite{Momma2008}. X-ray photoemission spectroscopy (XPS) measurements were performed at the AIPES (angle-integrated photoemission spectroscopy) beamline of Indus-1, RRCAT, Indore using the Omicron hemispherical energy analyzer (EA-125, Germany). The measurements were conducted at room temperature (300 K) using Al K$_\alpha$ X-ray radiation of energy 1486.6 eV. The total energy resolution was set at 0.8 eV. During the experiment, a base pressure of the order of $10^{-10}$ mbar was maintained in the experimental chamber. The Fermi level of a polycrystalline Au foil was used to calibrate the binding energy. Electrical and magnetic properties were measured using a Quantum Design physical property measurement system (PPMS) with a magnetic field of up to 9 Tesla within the temperature range of 2 and 350 K. AC-magnetic susceptibility was measured by ACMS option of PPMS. Heat capacity was measured in a PPMS by relaxation technique.

\subsection{Computational Details}

The first-principles DFT calculations were performed with a plane-wave basis set and projector-augmented wave (PAW) potentials, as implemented in the Vienna Ab initio Simulation Package (VASP)~\cite{PhysRevB.47.558, PhysRevB.54.11169}. The  exchange-correlation functional was approximated with the Perdew-Burke-Ernzerhof (PBE) formulation of the generalized gradient approximation (GGA). Considering the substantial electron-electron interactions beyond GGA at transition metal sites, we utilized the fully rotationally invariant +U approach~\cite{PhysRevB.48.16929, Anisimov_1997}. The Coulomb parameter (U) were set to 9 eV and 3 eV for Cu 3d and Ru 4d states, while the exchange parameters were set 0.9 eV and 0.8 eV, respectively, following the values used in the literature~\cite{Panda2016}. Variation of U value over the range of 1-2 eV was checked and found to keep the qualitative behaviour intact.  The valence electron configurations used for each element were La (5s\(^2\), 5p\(^6\), 5d\(^1\), 6s\(^2\)), Cu (4s\(^1\), 3d\(^1\)\(^0\)), Ru (4p\(^6\), 4d\(^7\), 5s\(^1\)), and O (2s\(^2\), 2p\(^4\)). For Sm, a fixed valence of Sm\(^{3+}\) was assumed, with the 4f electrons treated as core states, and its valence configuration was considered to be (5s\(^2\), 5p\(^6\), 5d\(^1\), 6s\(^2\)). All calculations used a converged Monkhorst-Pack k-point grid of 8 $\times$ 8 $\times$ 6 and a plane-wave energy cut-off of 600 eV. The internal atomic positions of crystal structures were optimized until the residual forces on each ion were reduced to less than 0.001 eV/$\AA$.

\textcolor{black}{To further investigate the possible effect of magnetism of Sm in Sm-bearing compounds, we carried out additional all-electron calculations using the full-potential linearized augmented plane-wave (FP LAPW) method, as implemented in the WIEN2k package~\cite{Blaha2019WIEN2kAA}. This approach enables a reliable description of the localized $4f$ states, which are not captured within the PAW framework used in VASP.  Spin-orbit coupling (SOC) was included self-consistently to properly account for the spin and orbital contributions of the Sm $4f$ states. The on-site Coulomb interaction $U = 6$ eV and a Hund’s exchange parameter $J = 0.6$ eV was applied on Sm site. To evaluate the net magnetic moment, we further examined the dependence on the Hubbard $U$ parameter and the choice of spin quantization axis for Sm.}

\section{Results and discussion }\label{3}

\subsection{Structural Characterization}

We employed room-temperature X-ray powder diffraction (XRD) to investigate the crystal structure of the as-grown compounds. Figs.~\ref{fig2}(a)-(c) show the powder XRD patterns overlapped with corresponding Rietveld refinements for LCRO, LSCRO, and SCRO. The Rietveld refinement confirms that all the diffraction peaks can be indexed to a monoclinic structure with a space group of P2$_{1}$/n (No. 14). These observations are supported by the Goldschmidt tolerance factor, which is a key indicator of structural stability in A$_{2}$BB$'$O$_{6}$ double perovskites. The tolerance factor is defined as \( t = \frac{r_A + r_O}{\sqrt{2}(<r_B> + r_O)} \), where \(r_A\) and \(r_O\) represent the ionic radii of the A-site cation and oxygen, respectively, and \( <r_B> \) is the average ionic radius of the B-site cations (B and B$'$). For an ideal cubic double perovskite structure, \( t = 1 \) corresponds to Fm-3m space group symmetry, and any reduction in symmetry leads to a decrease in the tolerance factor~\cite{Vasala2015}. A value of \( t < 0.97 \) typically results in the stabilization of a monoclinic structure~\cite{Manna2016}. In our case, with ionic radii reported by Shanon~\cite{Shannon1976} \(r_{\mathrm{Sm}^{3+}} = 1.24 \, \text{\AA}, r_{\mathrm{La}^{3+}} = 1.36 \, \text{\AA}, r_{\mathrm{Cu}^{2+}} = 0.73 \, \text{\AA}, r_{\mathrm{Ru}^{4+}} = 0.62 \, \text{\AA},\) and \(r_{\mathrm{O}^{2-}} = 1.40 \, \text{\AA}\), the calculated tolerance factors for LCRO and SCRO are 0.94053 and 0.8996, respectively. These values are consistent with the experimentally observed monoclinic crystal structure in both compounds. The results of the structural refinements are tabulated in the Table I of SM~\cite{SM}. The monoclinic structure in these systems arises from the alternating arrangement of \( \mathrm{CuO}_6 \) and \( \mathrm{RuO}_6 \) octahedra in a face-centered cubic framework, where Cu and Ru occupy the $B$ and $B'$ sites with 2c (0,$\frac{1}{2}$,0) and 2d ($\frac{1}{2}$,0,0) Wyckoff positions, respectively, while La\(^{3+}\) or Sm\(^{3+}\) ions are situated at the A-site with the 4e (\textit{x},\textit{y},\textit{z}) Wyckoff positions. The substitution of Sm\(^{3+}\) in the place of La\(^{3+}\) introduces systematic changes in the lattice parameters \textcolor{black}{[cf. Fig.~\ref{fig2}(e)]}, mainly due to the smaller ionic radius of Sm\(^{3+}\) (1.24\AA) compared to La\(^{3+}\) (1.36\AA), leading to a structural contraction. This in turn enhances the tilting of the \( \mathrm{CuO}_6 \) and \( \mathrm{RuO}_6 \) octahedra and thereby influencing the Cu–O–Ru bond geometry. The average octahedral tilt angle ($\phi$), calculated as \( \phi = \frac{180^\circ - \omega}{2} \) ($\omega $ is the Cu–O–Ru bond angle)~\cite{Kayser2013}, increases with increasing Sm doping concentration\textcolor{black}{, as can be seen from the Fig.~\ref{fig2}(f)}. The calculated values of \(\phi\) are approximately \textcolor{black}{\(14.6^\circ\)} for LCRO and \textcolor{black}{\(18.1^\circ\)} for SCRO. This increasing tilt angle changes the superexchange pathways between Cu 3d and Ru 4d orbitals, as well as the frustrated triangular environment of Ru\(^{4+}\) ion [cf. insets in Figs.~\ref{fig2}(a)-(c)]. As evident, the structural contraction induced by Sm doping leads to a distortion of the second-nearest-neighbor Ru–Ru triangular units, where the initially nearly equal triangle sides become increasingly unequal. \textcolor{black}{As categorized by P. W. Barnes \textit{et al.}~\cite{Barnes2006}, the Bragg reflections in double perovskites can be categorized into several groups. Among these, the R-point reflections indexed by all-odd Miller indices, are characteristic of the B-site cation ordering that defines the monoclinic P2$_{1}$/n symmetry, and are absent in the highly disordered orthorhombic phase. The intensity of these low-angle peaks is directly related to the degree of ordering. In our samples the first two R-point reflections are particularly sensitive to ordering as their intensities scale with the ordering parameter and drops with Cu and Ru site exchanges. We therefore used the intensities of these R-point peaks to quantify antisite disorder (ASD) (see Fig. \textcolor{black}{2} of SM~\cite{SM}). La$_2$CuRuO$_6$ and LaSmCuRuO$_6$ samples show an ASD (Cu/Ru intermixing) of approximately 10\%. The fully Sm-substituted end member, Sm$_2$CuRuO$_6$, exhibit ASD between 5\% and 10\%. For simplicity and best approximation in the subsequent analysis, we have considered a uniform 10\% ASD across the entire series.}

\textcolor{black}{X‑ray photoelectron spectroscopy (XPS) was employed to determine the oxidation states of both transition‑metal and rare‑earth ions, as these valence states are crucial for elucidating the nature of the electronic and magnetic interactions present in the samples. As discussed in the section V of the SM~\cite{SM}, the transition‑metal ions Cu and Ru adopt +2 and +4 oxidation states, respectively, while the rare‑earth La and Sm are stabilized in the +3 oxidation state.~\cite{Mullica1985,Wang2021,Shit2024,Hu2012,Guo2018,Fiermans1975,Qin2016}}

\subsection{DC Magnetization}

\begin{figure*}[ht]
\includegraphics[width=0.95\linewidth, clip=true]{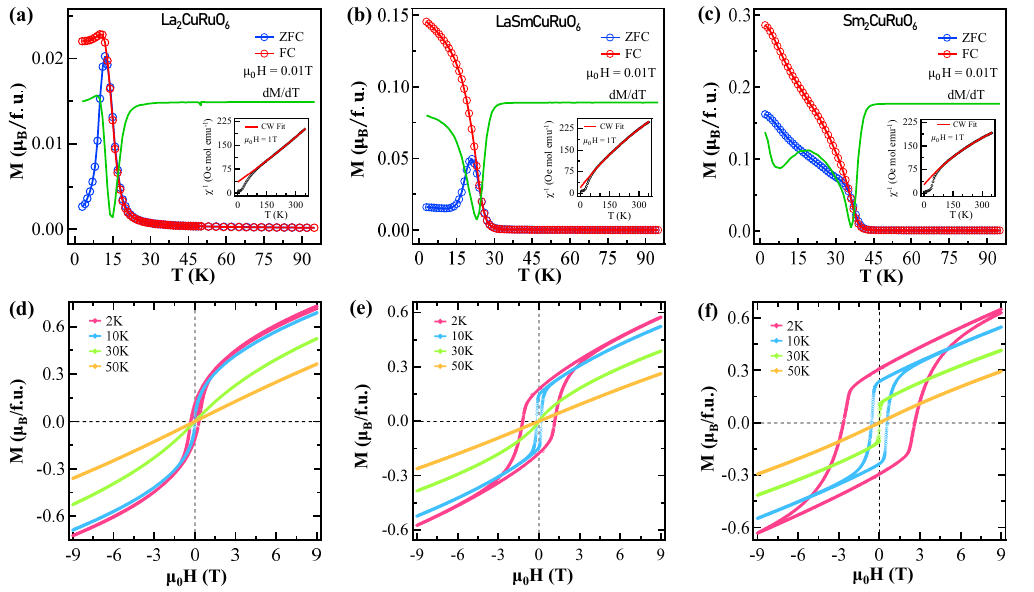}
\caption{(a)-(c) Present the variation of \textcolor{black}{magnetization with temperature along with first derivative of magnetization to estimate the transition temperature} for La$_2$CuRuO$_6$, LaSmCuRuO$_6$, and Sm$_2$CuRuO$_6$, respectively. The insets in (a)-(c) represent variation of \textcolor{black}{inverse dc magnetic susceptibility with temperature along with the Curie-Weiss law fitting}. (d)-(f) Presents the variation of magnetization with applied field for La$_2$CuRuO$_6$, LaSmCuRuO$_6$, and Sm$_2$CuRuO$_6$.}
\label{fig4}
\end{figure*}

Figs.~\ref{fig4}(a)–(c) display the temperature dependence of \textcolor{black}{magnetization along with its first derivative (dM/dT), which is used to determine the magnetic transition temperature} for LCRO, LSCRO, and SCRO. Magnetization measurements were carried out in both zero-field-cooled (ZFC) and field-cooled (FC) modes under an applied magnetic field of 10 mT. The insets in each panel show the temperature-dependent \textcolor{black}{inverse dc magnetic susceptibility along with the Curie-Weiss law fitting}. LCRO exhibits a typical paramagnetic behavior at high temperatures, while a distinct change in the slope of the magnetization curve at lower temperatures indicates a magnetic phase transition. From the minimum in the dM/dT curve, the transition temperature ($T_P$) is estimated to be approximately 15 K. In the paramagnetic region, the inverse susceptibility ($\chi^{-1}$) was fitted using the Curie-Weiss (CW) law, given by \(\chi_{CW} = \chi_0 + \frac{C}{T-\theta_{CW}}\), where $\chi_0$ represents the temperature-independent susceptibility, $C$ is the Curie constant, and $\theta_{CW}$ is the Curie-Weiss temperature~\cite{Naveen2018, Mugiraneza2022}. As the linear extrapolation of CW fitting intersects the negative temperature axis at 85K, the system is anticipated to have dominant antiferromagnetic (AFM) interactions. Notably, the magnitude of $\theta_{CW}$ is significantly larger than the transition temperature (\(T_P\)), highlighting the complexity of the system's magnetic behavior. This discrepancy arises because $\theta_{CW}$ reflects the overall strength of the magnetic interactions in the paramagnetic state, while the transition temperature is strongly influenced by the degree of magnetic frustration within the system~\cite{Kumar2012,Mugiraneza2022}.
In LSCRO, where Sm$^{3+}$ partially substitutes La$^{3+}$, $T_P$ increases to 23 K, while the Curie-Weiss temperature ($\theta_{CW}$) is found to be -22 K. Although the antiferromagnetic (AFM) interactions remain dominant, similar to LCRO, the rise in \(T_P\) suggests an enhancement in the magnetic transition temperature. In SCRO, where La$^{3+}$ is fully substituted by Sm$^{3+}$, $T_P$ increases further to 36 K accompanied by an additional low-temperature transition near 8 K. The $\theta_{CW}$ is estimated to be –36.2 K. These results suggest that the incorporation of Sm$^{3+}$ ions into LCRO drive the system towards dominant ferrimagnetic interactions from the dominant AFM interactions.

 \textcolor{black}{Further, below a specific temperature, the divergence between the ZFC and FC magnetization curves signifies the onset of intricate magnetic phenomena at low temperatures. The nature of this divergence is quite distinct across the three systems. The LCRO system exhibits a significant bifurcation between ZFC and FC curves, where ZFC curve shows a downturn below 15 K, with decreasing temperature and tending towards zero. This behaviour is consistent with earlier reports on LCRO characterizing its short-range magnetic correlations and glassy spin dynamics~\cite{Kumar2012,AnilKumar2017}. Similarly, in LSCRO also we noticed a downturn ZFC curve below 23 K. But instead of tending towards to zero magnetization with decreasing temperature the ZFC curve shows saturated magnetization below 15 K.  However, in SCRO the nature of ZFC-FC divergence is completely different than LCRO and LSCRO, as both FC and ZFC curves show  gradual increase in magnetization with decrease in temperature.  As discussed by P. A. Joy \textit{et~al.}~\cite{Joy1998}, in non-spin-glass systems, the degree of ZFC-FC irreversibility in the magnetization curves is intrinsically linked to the strength and temperature variation of the coercive field, \textcolor{black}{which may reflect} the intrinsic magnetic anisotropy of the system. To establish magnetic anisotropy unambiguously, one needs to perform the directional dependent magnetization measurements on the single crystals.}

Moreover, the $\chi_0$ value,  obtained by fitting with Curie-Weiss law, represents the sum of several intrinsic contributions, primarily the negative core diamagnetism ($\chi_{dia}$) arising from filled electron shells, the positive Van Vleck paramagnetism ($\chi_{VV}$), and the negligible positive Pauli paramagnetism ($\chi_{Pauli}$) for our insulating materials. The extracted values of $\chi_0$ for LCRO is $-7.74 \times 10^{-4}\text{emu} \text{ mol}^{-1} \text{Oe}^{-1}$, for LSCRO is $1.10 \times 10^{-3} \text{ emu} \text{ mol}^{-1} \text{Oe}^{-1}$, and for SCRO is $2.01 \times 10^{-3} \text{ emu} \text{ mol}^{-1} \text{Oe}^{-1}$. For LCRO, the core diamagnetic contributions from the closed-shell ions dominate, resulting in a small negative net $\chi_0$. Conversely, LSCRO and SCRO samples contain $\text{Sm}^{3+}$ ions, which are known to exhibit Van Vleck paramagnetic contribution due to the presence of low-lying excited states ($\text{J}=7/2$, $5/2$). This strong Van Vleck contribution from $\text{Sm}^{3+}$ outweighs the core diamagnetism in LSCRO and SCRO, leading to a net positive $\chi_0$. The observed increase in the magnitude of $\chi_0$ from LSCRO to SCRO directly correlates with the increasing concentration of the $\text{Sm}^{3+}$ ions. Typical core diamagnetism lies near $10^{-5} - 10^{-4} \text{ emu} \text{ mol}^{-1} \text{Oe}^{-1}$, while Van Vleck terms commonly reach $10^{-4} - 10^{-3} \text{ emu} \text{ mol}^{-1} \text{Oe}^{-1}$~\cite{Carlin1986, Guchhait2025}.
\textcolor{black}{The experimental effective magnetic moments ($\mu_{eff}^{exp}$) were determined from the Curie constants ($C$), which are 1.19 $emu~K~mol^{-1}$ for LCRO, 1.46 $emu~K~mol^{-1}$ for LSCRO, and 1.63 $emu~K~mol^{-1}$ for SCRO. These correspond to $\mu_{eff}^{exp}$ of 3.09 $\mu_B/f.u.$ for LCRO, 3.42 $\mu_B/f.u.$ for LSCRO, and 3.61 $\mu_B/f.u.$ for SCRO. To evaluate these experimental findings, the theoretical effective moments ($\mu_{eff}^{theo}$) were calculated based on the formal oxidation states of La$^{3+}$, Sm$^{3+}$, Cu$^{2+}$, and Ru$^{4+}$. The La$^{3+}$ ion is diamagnetic ($\mu_{La} = 0~\mu_B$), while for the transition metals, the spin-only contributions for Cu$^{2+}$ ($S = 1/2$, $\mu_{Cu} = 1.73~\mu_B$) and low-spin Ru$^{4+}$ ($S = 1$, $\mu_{Ru} = 2.83~\mu_{B}$) were utilized. For Sm$^{3+}$, the free-ion value derived from the Land\'{e} $g$-factor is notably low ($\mu_{Sm}^{theo} = 0.85~\mu_{B}$). However, as discussed above about the Van Vleck paramagnetism, an empirical effective value of $\mu_{Sm} \approx 1.50~\mu_B$ was chosen in accordance with related rare-earth double perovskite literature~\cite{Pradhan2021}. The resulting calculated theoretical values for LCRO, LSCRO, and SCRO are 3.32 $\mu_B/f.u.$, 3.64 $\mu_B/f.u.$, and 3.94 $\mu_B/f.u.$, respectively, which are in close agreement with the experimentally observed trends.}

Next, the isothermal magnetization as a function of applied magnetic field for LCRO, LSCRO, and SCRO is shown in Figs.~\ref{fig4}(d)-(f), illustrating the effect of Sm$^{3+}$ substitution. Across all compounds, the magnetization does not saturate even at the highest applied field of 9 T, which is consistent with the antiparallel alignment of Cu$^{2+}$ ($3d^9$) and Ru$^{4+}$ ($4d^4$) moments. Such behavior aligns with ferrimagnetism, where the magnetic moments on different sublattices are not completely cancelled, preventing from full saturation~\cite{Kumar2012}. \textcolor{black}{As can be seen from Fig.~\ref{fig4}(d), 2 K field dependant magnetization data for LCRO shows a typical sigmoid-S shape curve, representing the spin-glass state. Whereas, the 2 K $M(H)$ data of LSCRO and SCRO as shown in Figs.~\ref{fig4}(e) and ~\ref{fig4}(f),  respectively demonstrate significant coercive field ($H_C$) and remanent magnetization}. The coercive field at 2 K increases from 0.3 T for LCRO to 1.18 T for LSCRO and further to 2.63 T for SCRO (see Fig. \textcolor{black}{5} of SM~\cite{SM}), highlighting a stronger magnetic hysteresis with Sm substitution. Similarly, the remanent magnetization ($M_R$) increases from 0.11 $\mu_B$/f.u. for LCRO to 0.18 $\mu_B$/f.u. for LSCRO, and 0.30 $\mu_B$/f.u. for SCRO at 2 K (see Fig. \textcolor{black}{5} of SM~\cite{SM}). This enhancement in both $H_C$ and $M_R$ may be attributed to the induced lattice distortions due to smaller ionic radius of Sm$^{3+}$ compared to La$^{3+}$. These distortions enhance the magnetic anisotropy, making it difficult for the domain walls to move under the magnetic field, resulting in a higher coercive field~\cite{B.D.Cullity2008}.

\begin{figure}[ht]
	\includegraphics[width=0.95\linewidth, clip=true]{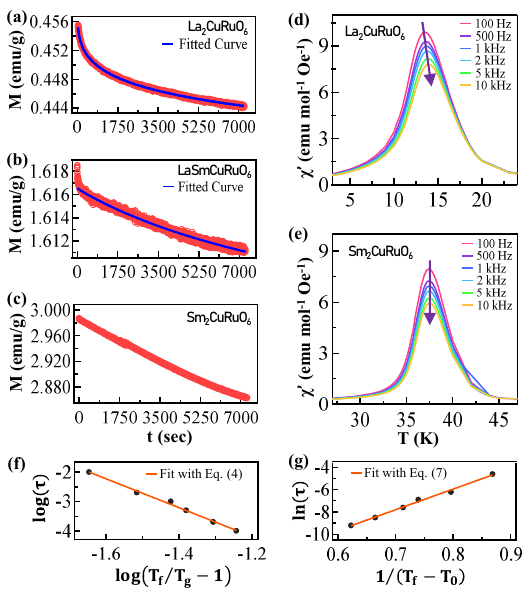}
	\caption{The relaxation of field-cooled dc magnetization as a function of time for (a) La$_{2}$CuRuO$_{6}$, (b) LaSmCuRuO$_{6}$, and (c) Sm$_{2}$CuRuO$_{6}$. Temperature dependence of real part ($\chi$$'$) of the ac susceptibility for (d) La$_{2}$CuRuO$_{6}$ and (e) Sm$_{2}$CuRuO$_{6}$.  \textcolor{black}{(f) Shows the fitting of relaxation time following the scaling law for La$_{2}$CuRuO$_{6}$. (g) Presents the fitting of relaxation time following Vogel-Fulcher law for La$_{2}$CuRuO$_{6}$.}}
	\label{fig5}
\end{figure}

The time-dependent relaxation of magnetization, measured in the field-cooled (FC) mode is depicted in Figs.~\ref{fig5}(a)-(c), provides key insights into the magnetic dynamics of LCRO, LSCRO, and SCRO. In this method, samples were cooled from 300 K to low temperature in an applied magnetic field of 500 Oe. After reaching low temperature, the field was removed, and following a wait time of 30 seconds, the time evolution of the magnetization was recorded over a two-hour period. The data was fitted using the stretched exponential function~\cite{Chamberlin1984,Bag2018,Sharma2019},
\begin{equation}
M(t) = M_R + M_g \exp\left[-\left(\frac{t}{\tau}\right)^\beta\right].
\end{equation}
Here, \(M_R\) is the remanent magnetization, \(M_g\) is the glassy magnetization component, \(\tau\) is the relaxation time, and \(\beta\) is the stretching exponent. The parameter \(\beta\) describes the distribution of relaxation times, where \(\beta = 1\) corresponds to a single relaxation time, and \(\beta < 1\) indicates a broad distribution of relaxation times, typically observed in disordered magnetic systems. In spin glass systems, \(\beta\) generally falls within the range of 0 to 1, reflecting the high degree of spin disorder and frustration~\cite{Chamberlin1984,Chu1994,Cardoso2003,Bag2018}. For LCRO, the fit yielded a stretching exponent \(\beta = 0.27\) and a relaxation time \(\tau \approx 66554\) s, which confirm the spin-glass behaviour. The small value of \(\beta\) suggests a disordered state with significant magnetic frustration~\cite{Kumar2012,AnilKumar2017}. In LSCRO, partial substitution of Sm$^{3+}$ increases \(\beta\) to 0.79 and decreases \(\tau\) to 49510 s.  This higher value of \(\beta\) suggests that the system is moving away from the glassy spin state towards more cooperative magnetic behaviour. Further, in SCRO, with full substitution of La$^{3+}$ by Sm$^{3+}$, $\beta$ exceeds 1. Thus, in SCRO spin-glass behaviour completely gets suppressed.

\subsection{AC Magnetization}
To further investigate the spin glass (SG) behaviour in these systems, and to understand the underlying nature of the transition, ac-susceptibility was measured at different frequencies ($\nu$). The real part of the ac-susceptibility ($\chi'$) as a function of temperature is shown in Figs.~\ref{fig5}(d) and ~\ref{fig5}(e) for LCRO and SCRO, respectively. Fig.~\ref{fig5}(d) shows the $\chi'$(T) of LCRO at $\nu = 100 Hz$ exhibiting a pronounced anomaly at around 13.5 K. The peak position shifts towards higher temperatures and the height of peak decreases with increasing frequency. This is a characteristic feature of a spin glass (SG) transition~\cite{Kumar2012,Naveen2018,Bag2018}.
The relative shift in freezing temperature per decade of frequency, often referred to as the Mydosh parameter ($\delta T_f$), is a commonly used metric for comparing different spin glass systems. This parameter can be determined using the equation~\cite{Mulder1982,Mydosh2014,Mydosh2015},
\begin{equation}
    \delta T_f = \frac{\Delta T_f}{T_f\Delta(log_{10}\nu)}
\end{equation}
where \(\Delta T_f = (T_f)_{\nu_1} - (T_f)_{\nu_2}\) and \(\Delta(log_{10}\nu) = log_{10}(\nu_1)-log_{10}(\nu_2)\). To calculate $\delta T_f$, we employed two outermost frequencies $\nu_1$ = 100 Hz and $\nu_2$ = 10 kHz. For LCRO, we estimate $\delta T_f \approx 0.017$. The obtained value places the system in an intermediate regime. It is roughly ten times larger than what is typically observed in canonical spin glass systems~\cite{Mulder1982,Mydosh2014,Mydosh2015}. Notably, this value lies within the range commonly associated with cluster spin glass systems~\cite{Malinowski2011,Mydosh2014,Bag2018}. In conventional magnetic systems, where the interaction between magnetic ions is strong, a significantly large frequency is required to produce a noticeable shift in ac-susceptibility. However, in a magnetic cluster system, the interactions between clusters are weak, leading to a stronger frequency sensitivity.

To know the spin dynamics, frequency-dependent freezing temperature ($T_f$) is obtained from the real part of ac-susceptibility. In SG systems, the frequency dependence of $T_f$ can be described by a conventional power law of the critical-slowing-down model, given by the dynamic scaling theory~\cite{Binder1986,Mydosh2014}. The relaxation time $\tau$, corresponding to the measured frequency (\(\tau = 1/\nu\)), describes the dynamical fluctuation time scale as,
\begin{equation}
\tau = \tau_0\left(\frac{T_f - T_g}{T_g}\right)^{z\nu'}
\end{equation}
where $\tau_0$ is the relaxation time of a single spin-flip of the fluctuating entities, $T_g$ is the static freezing temperature as $\nu$ tends to zero, and $z\nu'$ is the dynamic critical exponent. To better fit the data, the power law can be reformulated as,
\begin{equation}
\log_{10} \tau = \log_{10} \tau^* - z \nu' \log_{10} \left( \frac{T_f}{T_g} - 1 \right).
\end{equation}
\textcolor{black}{Fig.~\ref{fig5}(f)} presents the plot of \(\log_{10}(\tau)\) vs. \(\log_{10}\left({T_f}/{T_g} - 1\right)\) with \(T_g\) fixed at \(13.22 \text{K}\). The parameters derived from the optimal fit of the data are \(\tau_0 \approx 7.63 \times 10^{-10} \text{s}\) and \(z\nu' \approx 4.93\). The dynamic scaling analysis reveals a compelling divergence of the relaxation time at a finite transition temperature, confirming the phase transition from paramagnetic (PM) to spin-glass (SG) behaviour in LCRO. The parameters \(\tau_0\) and \(z\nu'\) are particularly noteworthy as they are believed to offer profound insights into the SG dynamics. In conventional SG systems, the \(z\nu'\) values typically lie within 4 and 12, while \(\tau_0\) spans between \(10^{-10} \text{s}\) and \(10^{-13} \text{s}\). Intriguingly, for canonical SG and cluster SG systems, the characteristic range of \(\tau_0\) varies from \(10^{-12} \text{s}\) to \(10^{-13} \text{s}\) and \(10^{-7} \text{s}\) to \(10^{-10} \text{s}\), respectively~\cite{Lago2012,Malinowski2011,Mydosh2014,Bag2018,Naveen2018}. Therefore, our findings of \(\tau_0\) and \(z\nu'\) suggest a typical cluster SG system. 

The evidence for interacting clusters is also reflected in the inability of the Arrhenius law to describe the frequency-dependent \( T_f \) data (not shown). Typically, the Arrhenius law applies to systems with non-interacting or weakly interacting magnetic moments, which is expressed as~\cite{Binder1986,Bag2018},

\begin{equation}
\tau = \tau_0 \exp \left( \frac{E_a}{k_B T_f} \right)
\end{equation}

where \( \tau_0 \) is the characteristic relaxation time, and \( E_a / k_B \) is the average activation energy of the relaxation barrier. The activation energy represents the barrier separating metastable states, and the Arrhenius law describes the characteristic timescale required to overcome these barriers via thermal activation. We attempted to determine \( \tau_0 \) and \( E_a / k_B \) from a linear fit of \( \ln(\tau) \) vs. \( 1/T_f \), but the results were unphysical (\( \tau_0 \approx 3.16 \times 10^{-63} \, \mathrm{s} \) and \( E_a / k_B \approx 1879 \pm 122 \, \mathrm{K} \)). This further reinforces that the spin dynamics in this system are not governed by simple single spin-flip processes, instead involve the complex magnetic interactions between the clusters~\cite{Bag2018,Naveen2018}.

A widely used approach to describe the spin-glass freezing dynamics is the Vogel-Fulcher (VF) law, a phenomenological model that captures the impact of spin interactions on the freezing process. VF law characterizes the frequency dependence of the freezing temperature \( T_f \) as~\cite{Souletie1985,Mydosh2014},

\begin{equation}
\tau = \tau_0 \exp\left( \frac{E_a}{k_B(T_f - T_0)} \right)
\end{equation}

where \( T_0 \) is the Vogel-Fulcher temperature, which is generally interpreted as a measure of the strength of interaction between the spins. Further, the VF equation can be linearized as,

\begin{equation}
\ln \tau = \ln \tau_0 + \frac{E_a / k_B}{T_f - T_0}.
\end{equation}

Using Eq. (7), we analyzed the experimental data over the accessible frequency range and found that the VF law accurately describes the behaviour of \( T_f \). The plot of \( \ln \tau \) vs. \( 1/(T_f - T_0) \), shown in the \textcolor{black}{Fig.~\ref{fig5}(g)}, demonstrates fit to this equation with $T_0 \approx 12.37\, \mathrm{K}$. The values of $E_a / k_B \approx 18.5 \, \mathrm{K}$ and $\tau_0 \approx 9.6 \times 10^{-10} \, \mathrm{s}$ are found from the slope and intercept on the y-aixs of the fitting. The non-zero \( T_0 \) value and the strong agreement with the VF law indicate that finite interactions exist among spins, resulting in the formation of clusters. From the above analysis, it is evident that the variation in relaxation time, \( \tau \), within the experimental frequency range can be equally well described by both the power law and the Vogel-Fulcher (VF) law. The characteristic relaxation time \( \tau_0 \), obtained from the power law, is approximately an order of magnitude smaller than the value, as derived from the VF law~\cite{Bag2018,Naveen2018}. Nonetheless, the values of \( \tau_0 \) from both fits fall within the expected range for typical cluster spin-glass systems~\cite{Bag2018,Naveen2018}.

Fig.~\ref{fig5}(e) displays the ac-susceptibility data for SCRO. In SCRO, the peak position remains unchanged across a broad frequency range. This frequency independence of peak position suggests that the observed transition is not a characteristic of spin$-$glass phase. Instead, it points towards a magnetically \textcolor{black}{cooperative} state, where the dynamics are governed by collective excitations rather than the slow, frustrated dynamics as seen in spin glasses.

\subsection{Heat Capacity}

\begin{figure}[]
\includegraphics[width=0.9\linewidth, clip=true]{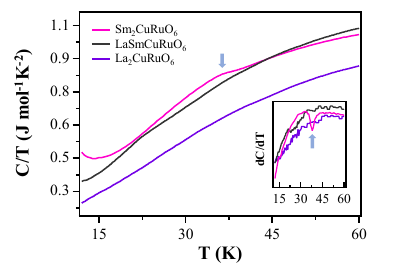}
	\caption{The $C/T$ versus T plot near the magnetic anomaly temperature region is shown for three systems. The inset shows the derivative of C with respect to T for three systems.}
	\label{fig-CT}
\end{figure}

To understand the degree of magnetic ordering in these systems,  we measured the temperature dependence of the zero-field heat capacity, $C(T)$ for LCRO, LSCRO, and SCRO, respectively. For a better visualization, we have plotted $C/T$ as a function of T as shown in Fig.~\ref{fig-CT} for all three systems to highlight the anomaly in specific heat. We can clearly notice that SCRO shows a kink at around 37 K where the magnetization data [$M(T)$] show the transition and no such kink is observed from LCRO and LSCRO. We also present the derivative of C(T) with respect to T for all three systems near magnetic anomaly temperature region, as shown in the inset of Fig.~\ref{fig-CT} to exactly pinpoint the transition temperature in SCRO. \textcolor{black}{To complement these findings, we analyzed the heat capacity data as outlined in the section VII of SM~\cite{SM} by modeling the lattice contribution with a Debye-Einstein term~\cite{Datta2023,Guchhait2025}, accounting for Sm$^{3+}$ crystal-field splitting~\cite{Lopez2002,Harikrishnan2008,Mazumdar2021, Patel2025}, and applying the Schottky equation to the low-temperature regime~\cite{Fertman2009,Mazumdar2021}. This suggest a more cooperative magnetic interactions in SCRO than the short range magnetic correlation present in LCRO.}


\subsection{Mechanism for Tunable Magnetic State in La$_{2-x}$Sm$_x$CuRuO$_6$}

Further, to understand the microscopic origin of the contrasting
magnetic behavior of La\(_2\)CuRuO\(_6\) and Sm\(_2\)CuRuO\(_6\) we carried out
DFT calculations. \textcolor{black}{There exists only few reports of first-principles calculations for La$_2$CuRuO$_6$\cite{Panda2016, Kumar2012} which
confirmed the existence of nearest-neighbor Cu-O-Ru and Ru-Ru next nearest-neighbor interactions, highlighting the strong antiferromagnetic behavior responsible for the magnetic frustration observed experimentally in the system. However, to the best of our knowledge, no systematic study has been carried
out on Sm$_2$CuRuO$_6$ or on the trend as La is replaced by an isovalent Sm atom. Here, we present the calculations of two different cases (i) without considering Sm magnetic moment and (ii) considering Sm magnetic moment.}

In first step towards our study, the crystal structures were theoretically
optimized keeping the lattice constants and symmetries intact. While XRD provides accurate atomic positions of heavy elements, for relatively light elements like O,
the estimated positions may not be very accurate. In accordance, with this expectation, some of the O positions in the theoretically optimized structures show large deviation of 20-30$\%$, proving the structural optimization to be crucial. The DFT optimized structures were found to reproduce the experimentally observed insulating ground state.

\begin{figure}[]
	\includegraphics[width=1\linewidth, clip=true]{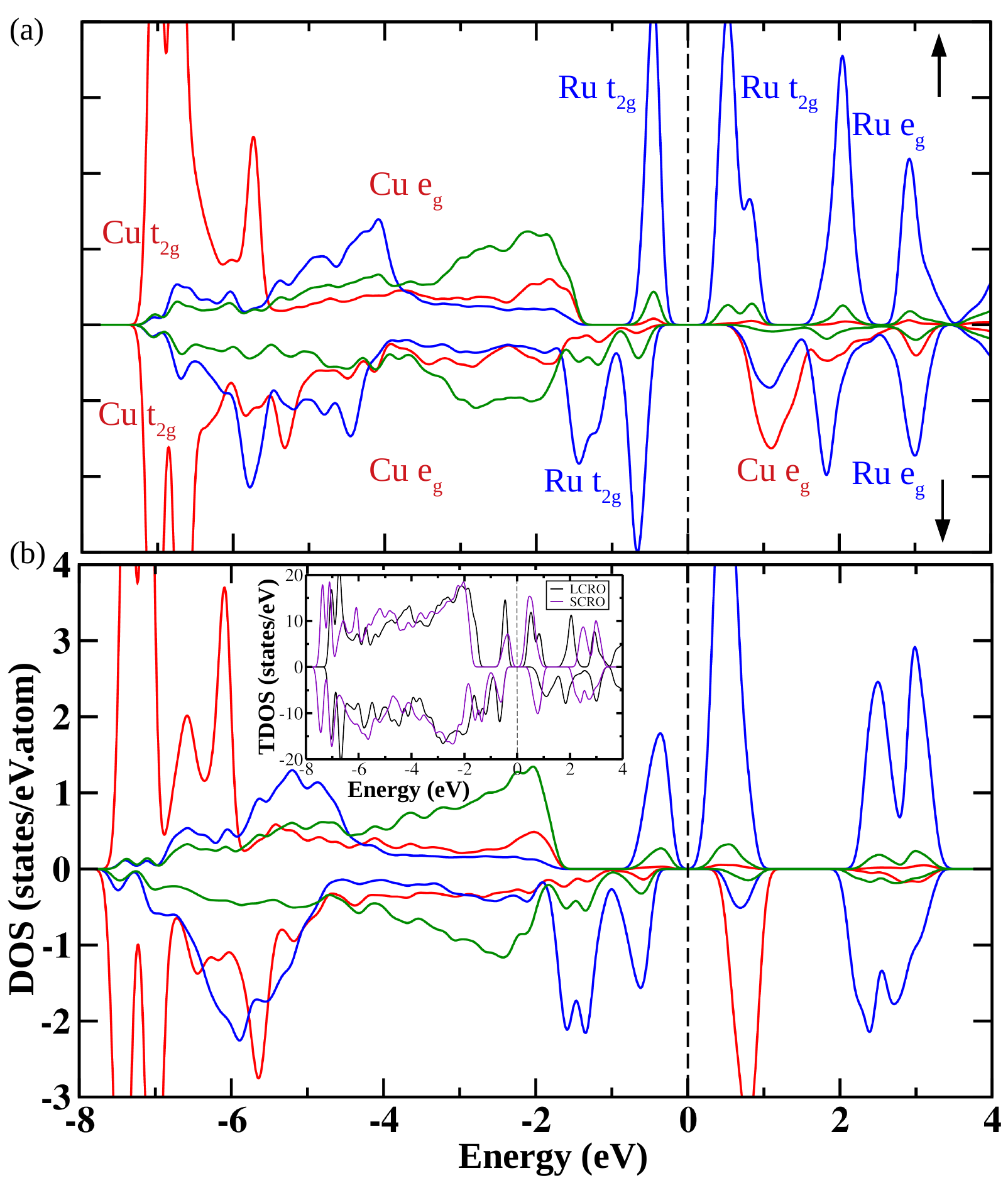}
	\caption{(a) The GGA+U density of states of LCRO, projected to Cu-$d$ (red), Ru-$d$ (blue), and O-$p$ (green) states.  The zero of the energy is set Fermi level. The positive (shown as the upward arrow ($\uparrow$))
    and negative (shown as the downward arrow ($\downarrow$)) density of states denote the majority and minority spin channel. (b) The GGA+U density of states of SCRO, projected to Cu-$d$ (red), Ru-$d$ (blue), and O-$p$ (green) states. The comparison of total density of states (TDOS) between LCRO (black) and SCRO (violet) are shown as inset.}
	\label{fig-theory}
\end{figure}

Fig.~\ref{fig-theory}(a) shows the GGA+U spin-polarized
density of states (DOS) of LCRO projected to Cu $d$, Ru $d$ and O $p$ states.
The Cu as well as Ru $d$ states are crystal field split into $t_{2g}$ and
$e_g$ as well as exchange split. The Jahn-Teller active Cu$^{2+}$ $d^9$
ion in the compressed CuO$_6$ octahedral environment has $d_{z^{2}}$ state half filled, all other states being filled.
The inclusion of large $U$ at Cu site opens up a gap between filled Cu $d_{z^{2}}$ up spin state and empty Cu $d_{z^{2}}$ down spin state. Ru in its nominal +4 oxidation state, on other hand, stabilizes in
low-spin 4$d^4$ ($t_{2g}^4$) configuration. The Ru $t_{2g}$states are mostly filled in the down spin channel and partially filled in the up spin channel.
The inclusion of $U$ at Ru site opens up a gap of $\sim$ 0.35 eV between
the Ru $t_{2g}$ states in up spin channel, which are non degenerate
due to deviation from perfect octahedral symmetry of
RuO$_6$. In the down spin channel the empty Cu $d_{z^{2}}$ states
remain separated from Ru $t_{2g}$ states by a larger gap of $\sim$ 0.8 eV.
The Cu majority and Ru majority
spin states are oppositely aligned, suggestive of a ferrimagnetic ground
state, as observed experimentally. The gross features of DOS for SCRO, shown in
Fig.~\ref{fig-theory}(b) are the same as LCRO. The band gap in SCRO
is found to reduce from 0.35 eV in LCRO to 0.2 eV (cf. inset in Fig.~\ref{fig-theory}(b)), in good agreement with the trend observed experimentally \textcolor{black}{(cf. section VIII of SM~\cite{SM})}.
This reduction happens to the increased band-width of Ru $t_{2g}$ states due to contraction of the RuO$_6$ octahedra in SCRO.

The calculated magnetic moments support the
findings from analysis of DOS. The Cu and Ru magnetic moments
turn out to be 0.76 (0.76) $\mu_B$ and -1.38 (-1.44) $\mu_B$ for
LCRO (SCRO). The deviation from Cu and Ru \textcolor{black}{spin-only} nominal magnetic moment of
1 $\mu_B$ and 2 $\mu_B$ are explained by large Cu-O and Ru-O hybridization
with missing moments at O sites and at interstitials. The total moment
is found to be integer, -\(1.0 \, \mu_B \)/f.u. for both compounds, in
agreement with their insulating nature. Calculations including spin-orbit coupling (SOC) show that the orbital moment at the Ru site is significant, ranging from approximately -$0.10$ to -$0.13~\mu_B$, approximately three times greater than that of the Cu ions ($\sim$ 0.04 $\mu_B$).

\begin{figure}[]
	\includegraphics[width=1\linewidth, clip=true]{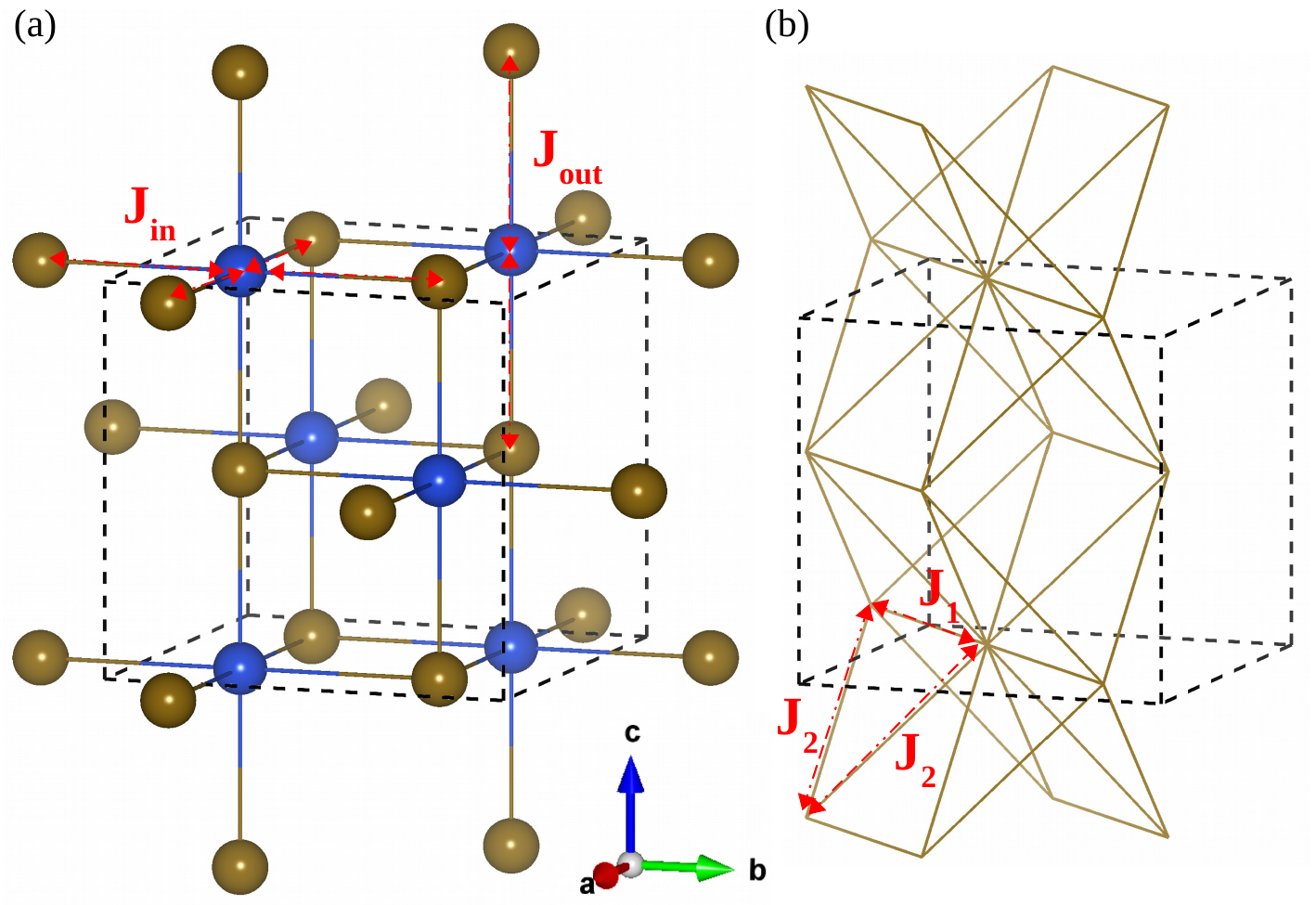}
	\caption{(a) The nearest neighbour in-plane and out-of-plane Cu-Ru magnetic exchanges, J$_{in}$ and J$_{out}$. Only Cu (blue) and Ru (dark brown) atoms are shown for clarity. (b) Second nearest-neighbor Ru-Ru interactions, J$_1$ and J$_2$,
    in the Ru network (shown as wireframes, where each connecting point represents a Ru ion) of a double perovskite structure.}
	\label{J-calc}
\end{figure}

To investigate the Cu-Ru and Ru-Ru magnetic interactions, we calculated GGA+U total energies of different spin configurations. The Cu-O-Ru superexchange angle turns to be different
by $\sim$ 1-2 $^{o}$ making the in-plane and out of plane Cu-Ru exchanges
different. The two arms of the triangle formed by Ru-Ru bonds, as shown in inset of Fig.~\ref{fig2}, are nearly equal, with distinctly different third arm for
Sm compound. We thus considered two different Cu-Ru (J$_{in}$ and J$_{out}$) exchanges,
and two different Ru-Ru (J$_1$, J$_2$) exchanges, as shown in Fig.~\ref{J-calc}.
To estimate these four exchanges, five different spin arrangements of Cu and
Ru spins were considered including the parallel alignment of Cu and Ru spins.
Total energy differences were mapped onto a Heisenberg Hamiltonian: \(\hat{H} = -\sum_{i \neq j} J_{ij} \vec{S}_i \cdot \vec{S}_j\), where \(i\) and \(j\) are the positions of the magnetic ions, i.e., Cu and Ru, and \(\vec{S}_i\) and \(\vec{S}_j\) are the spin vectors at sites \(i\) and \(j\), respectively. The magnitude and direction of these spins determine the nature of the magnetic interactions between neighbouring ions, positive values implying ferromagnetic coupling and negative values implying antiferromagnetic coupling. The estimated J's are listed in Table~\ref{tab:exchange_parameters}.
Firstly, we find all J's to be negative. For a perfectly cubic double perovskite structure, with 180\textdegree \ Cu-O-Ru bond angle, no hopping between Cu d$_{z^2}$ and Ru $t_{2g}$ states can take place due to orthogonality of the orbitals. However, the crystal structures of LCRO and SCRO exhibit a significant amount of octahedral rotations, leading to large deviation Cu–O–Ru bond angles by more than 30\textdegree from 180\textdegree. These distortions create finite super-exchange AFM interaction
between the Cu d$_{z^2}$ and Ru $t_{2g}$ states.
Secondly we find presence of a substantially large second-neighbour Ru-Ru interaction arising from direct overlap of extended 4d orbitals of Ru. The corresponding Cu-Cu interaction, as checked independently, turned out to be small. Ru-Ru interactions
between three unpaired low-spin 4t$_{2g}^4$ turn to be also AFM, as Ru $e_g$ states
have large crystal field splitting from Ru $t_{2g}$ states.
Nearest neighbor AFM Cu-Ru interaction does not yield frustration. However, the
second neighbor Ru-Ru interaction can give rise to frustration due to the triangular geometry of B$'$ network of the double perovskite structure. Considering the theoretical optimized structures, for La compound, the Ru-Ru bondlengths show a maximum deviation of 0.003 $\AA$, while Sm compound this shows more than order of magnitude increase to 0.076 $\AA$. This is reflected in the computed J's. The J$_1$ and J$_2$ differ by only 0.02 meV for La compound and differ by a large amount of 0.67 meV for Sm compound. The dominant frustration effect in La compound is thus lifted for Sm compound, enabled by
structural distortion due to introduction of smaller cation Sm$^{3+}$ with ionic
radius of 1.24 $\AA$, in place of La$^{3+}$ having ionic radius of 1.36 $\AA$.
The competing AFM exchanges in LCRO thus is responsible for the glassy characteristics observed experimentally, which weakens significantly in SCRO, removing such signatures.

The above analysis was carried without considering the possible magnetism  of Sm. However Sm$^{3+}$ may possess magnetic moment. The theoretical magnetic moment
of Sm$^{3+}$, calculated using Lande’ g factor and total angular moment quantum number (J=5/2), is $\sim$ 0.85 $\mu_B$. However, this can be strongly influenced by
the crystal-field and surrounding effects and can even be fully quenched, as in the case of SmN~\cite{PhysRevB.78.174406}. Calculation of the spin and orbital moment of Sm in Sm$_2$CuRuO$_6$ and LaSmCuRuO$_6$ compounds, within the all electron method of full-potential LAPW resulted in spin moment of $\sim$ 4.98 $\mu_B$, while the orbital
moment turned out to be 2.79-2.27 $\mu_B$, oppositely aligned to the spin moment, depending on the choice of U parameter on Sm and the spin quantization axis (see Table \textcolor{black}{V} of SM~\cite{SM}).
Thus, a net moment of 2.10-2.72 $\mu_B$ develops at the Sm site, making the Sm ion magnetic. Total energy calculations considering different orientations of
Sm, Cu, and Ru magnetic moments further reveal that Sm moment points parallel to Ru, and antiparallel to Cu.
\textcolor{black}{As presented in SM~\cite{SM} (see Table \textcolor{black}{VI} of SM), the strengths of Cu-Sm and Ru-Sm exchange interactions are found to be about 10$\%$ of the
dominant Ru-Ru interaction, while Sm-Sm interaction is found to  be further smaller by a factor of 5-6. Nevertheless, inclusion of these additional
contributions in Sm-bearing compound in the
highly sensitive, magnetically frustrated system
of Sm-free compound, results in relief of frustration. Thus, the lattice effect coupled with effect of Sm magnetism plays an important role in progressive suppression of frustration effect as La is replaced by Sm.}

\begin{table}[]
\caption{Computed Cu-Ru and Ru-Ru magnetic exchange interactions.
Negative values denote antiferromagnetic coupling.}
\centering
\renewcommand{\arraystretch}{2.5}
\begin{tabular}{|@{}l|cc|c|@{}|}
\toprule
\multirow{2}{*}{\textbf{Compound}} & \multicolumn{2}{c|}{\textbf{Cu-Ru}} & \multicolumn{1}{c|}{\textbf{Ru-Ru}} \\
\cmidrule{2-4}
 & \boldmath$J_{\text{in}}$ \textbf{(meV)} & \boldmath$J_{\text{out}}$ \textbf{(meV)} & \boldmath$J_{1}$, \boldmath$J_{2}$ \textbf{(meV)} \\
\midrule
LCRO              & -2.49                                  & -2.46                                  & -5.92, -5.94                                  \\
SCRO              & -3.81                                  & -3.56                                  & -6.04, -5.63                                  \\
\bottomrule
\end{tabular}
\label{tab:exchange_parameters}
\end{table}

\section{Conclusions}\label{4}

\textcolor{black}{In conclusion, our systematic experimental investigation corroborated with first-principles study of La$_2$CuRuO$_6$, LaSmCuRuO$_6$, and Sm$_2$CuRuO$_6$ highlights the critical role of rare-earth substitution in modulating the structural, magnetic, and electronic properties of copper-ruthenate double perovskites. The progressive replacement of La$^{3+}$ with Sm$^{3+}$ ions induces significant lattice distortions, which, in turn, weakens the competition between AFM  magnetic interactions and lead to the lifting of frustration in Sm$_2$CuRuO$_6$. This works hand-in-hand with \textcolor{black}{the effect of Sm magnetism that introduces additional  AFM Sm-Cu and FM Sm-Ru magnetic exchanges.} Thus, the spin-glass behaviour observed in La$_2$CuRuO$_6$ gets suppressed in Sm$_2$CuRuO$_6$, demonstrating the sensitivity of magnetic properties in these materials to the A-site cation. Furthermore, the electrical resistivity measurements reveal a variable-range hopping~\cite{Mott1969,Mott1979} domination at low temperatures (cf. section \textcolor{black}{VIII} of SM~\cite{SM}), indicative of strong electron localization in these systems.  These findings shed light on the complex interplay of structure and magnetism in Cu-Ru double perovskites, and open up an avenue for tuning the magnetic properties through rare-earth substitution.}

\textcolor{black}{It is worth mentioning, additionally, antisite defects (where Cu and Ru ions exchange their positions) may also contribute to frustration effect. In the synthesized compounds, the antisite disorder is estimated to be around 5-10\%, which is not large, but their presence may enhance magnetic frustration and reduce the overall magnetic ordering temperature.}

\textcolor{black}{Finally, to further pinpoint the influence of substitution of La$^{3+}$ by a smaller cation Sm$^{3+}$ on the Ru network of the double perovskite structure, we theoretically examined the crystal structure by substituting magnetic Cu$^{2+}$ with the non-magnetic ion Zn$^{2+}$ in both La and Sm compounds. The theoretically optimized Zn substituted structures show the Ru-Ru bond length differ by 0.004~\AA~ in La compound and 0.078~\AA~ in Sm compound, compared to a difference of 0.003~\AA~ and 0.076~\AA~, respectively, for the Cu based structure. This finding motivates synthesis and characterization of Zn compounds, which will be taken up in future.}

\section{Acknowledgement}\label{5}

S.G. acknowledges University Grants Commission (UGC), India for the Ph.D. fellowship. T.S-D acknowledges the J.C.Bose National Fellowship (grant no. JCB/2020/000004) for support. The authors acknowledge the financial support through the India-Russian joint research project (DST-RSF), having the grant number DST/INT/RUS/RSF/P-53/2021 (G). The authors acknowledge RSCF grant No. 25-12-00028 for measurements of ac-susceptibility. The authors would like to acknowledge Mr. Sharad Karwal for his valuable assistance during the XPS measurements. \textcolor{black}{Heat capacity measurements were facilitated under the MoU between CSIR-National Physical Laboratory and S N Bose National Center for Basic Sciences. The authors would like to acknowledge Dr. Pallavi Kushwaha for her help in the heat capacity measurements.} This research has made use of the Technical Research Centre (TRC) Instrument Facilities of the S. N. Bose National Centre for Basic Sciences, established under the TRC project of the Department of Science and Technology Government of India.

\newpage

\section*{Supplementary Information}
\textcolor{black}{\section{Frustration in Double Perovskite Structure}}

\begin{figure}[ht]
	\includegraphics[width=1\linewidth, clip=true]{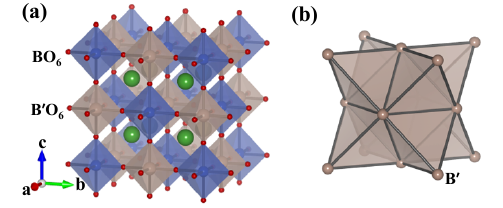}
	\caption{\textcolor{black}{(a) Schematic presentation of cubic double perovskite structure of $A_2BB'O_6$ showing $BO_6$ and $B'O_6$ octahedra forming interpenetrating fcc lattice. (b) $B'$ ions form a frustrated fcc sublattice made of edge-sharing tetrahedra.}}
	\label{fig-DP}
\end{figure}

\section{Estimation of Antisite Disorder}

\begin{figure}
    \centering
    \includegraphics[width=0.5\textwidth]{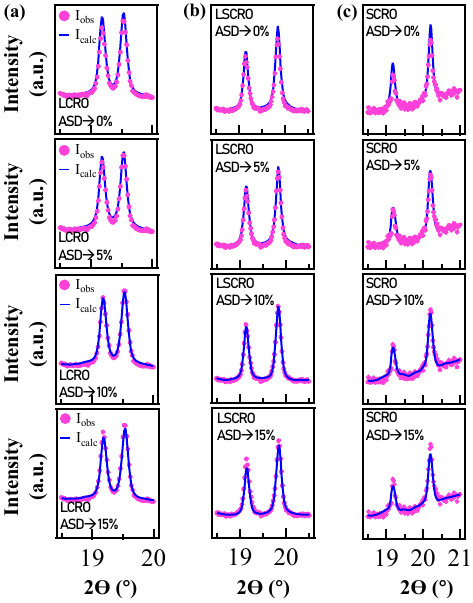}
    \caption{Rietveld refinement using the monoclinic $P2_1/n$ structure by varying the antistite disorder for (a)La$_2$CuRuO$_6$ (LCRO), (b)LaSmCuRuO$_6$ (LSCRO), and (c)Sm$_2$CuRuO$_6$ (SCRO).}
    \label{fig-ASD}
\end{figure}

We analyzed the influence of antisite disorder (ASD) as the best fitting of the low-angle double peak structure (R-point reflections~\cite{Barnes2006}) in the experimental XRD pattern, by systematically varying the site occupancies of Cu and Ru at the 2$c$ and 2$d$ Wyckoff positions within the monoclinic $P2_1/n$ symmetry, as can be seen from Fig.~\ref{fig-ASD}.

\section{Rietveld Refinement Results}

The result of the Rietveld refinement of XRD data obtained at room temperature are detailed in the following Table~\ref{tab-struc}.

\begin{table*}[]
\caption{\label{tab-struc} Structural parameters and positional parameters after the Rietveld refinement of XRD data obtained at room temperature.}
\begin{ruledtabular}
\begin{tabular}{lccc}
\textrm{Compound}& \textrm{La$_{2}$CuRuO$_{6}$}& \textrm{LaSmCuRuO$_{6}$}& \textrm{Sm$_{2}$CuRuO$_{6}$}\\
\textrm{Space group}& \textrm{P2$_{1}$/n (14)}& \textrm{P2$_{1}$/n (14)}& \textrm{P2$_{1}$/n (14)}\\
\colrule
\multicolumn{4}{l}{\textit{Cell Parameters}}\\
a (\AA) &5.584(3) &5.499(3) &5.400(6) \\
b (\AA) &5.750(7) &5.816(2) &5.849(7) \\
c (\AA) &7.749(4) &7.644(1) &7.527(1) \\
\(\beta (^\circ)\) &89.951(6) &89.989(7) &90.069(2) \\
V (\AA³) &248.86(4) &244.49(3) &237.80(6) \\

\multicolumn{4}{l}{\textit{Atomic Positions RE 4e (x, y, z)}}\\
x &0.5020(3) &0.5097(1) &0.5119(7) \\
y &0.5517(4) &0.5623(5) &0.5715(5) \\
z &0.2499(1) &0.2490(2) &0.2499(5) \\

\multicolumn{4}{l}{\textit{Cu 2c(0, 1/2, 0)}}\\
\multicolumn{4}{l}{\textit{Ru 2d (1/2, 0, 0)}}\\

\multicolumn{4}{l}{\textit{O1 4e (x, y, z)}}\\
x &0.2061(4) &0.1784(8) &0.1727(1) \\
y &0.1832(8) &0.2013(7) &0.1764(9) \\
z &-0.0306(3) &-0.0347(4) &-0.0603(6) \\
\multicolumn{4}{l}{\textit{O2 4e (x, y, z)}}\\
x &0.3006(5) &0.3284(1) &0.3386(4) \\
y &0.6762(4) &0.7051(9) &0.6817(8) \\
z &-0.0492(7) &-0.0453(1) &-0.0478(8) \\
\multicolumn{4}{l}{\textit{O3 4e (x, y, z)}}\\
x &0.4250(8) &0.4000(3) &0.4204(6) \\
y &0.0024(3) &-0.0295(7) &-0.0267(9) \\
z &0.2635(2) &0.2458(4) &0.2659(5) \\

\multicolumn{4}{l}{\textit{Bond Lengths}}\\
Cu-O1 (\AA) &2.1675(9) &2.0126(3) &2.1584(3) \\
Cu-O2 (\AA) &1.9976(7) &2.1922(1) &2.1464(2) \\
Cu-O3 (\AA) &1.8801(7) &2.0165(1) &1.8196(1) \\
Ru-O1 (\AA) &1.9649(0) &2.1374(3) &2.0964(2) \\
Ru-O2 (\AA) &2.0842(4) &1.9875(7) &2.0865(9) \\
Ru-O3 (\AA) &2.2027(6) &1.9654(1) &2.0539(3) \\

\multicolumn{4}{l}{\textit{Bond Angles}}\\
Cu-O1-Ru ($^\circ$) &151.763(1) &149.304(7) &138.644(6) \\
Cu-O2-Ru ($^\circ$) &145.133(7) &146.436(5) &140.242(1) \\
Cu-O3-Ru ($^\circ$) &155.548(6) &146.441(8) &152.582(1) \\

\({<}Cu-O-Ru{>} (^\circ)\) &150.82 &147.39 &143.82 \\
\({<}\phi{>} (^\circ)\) &14.59 &16.30 &18.09 \\

G.O.F &3.63 &4.55 &3.91 \\

\end{tabular}
\end{ruledtabular}
\end{table*}

\section{Rietveld Refinement by Changing Occupancy}
\begin{figure*}[]
    \centering
    \includegraphics[width=0.8\textwidth]{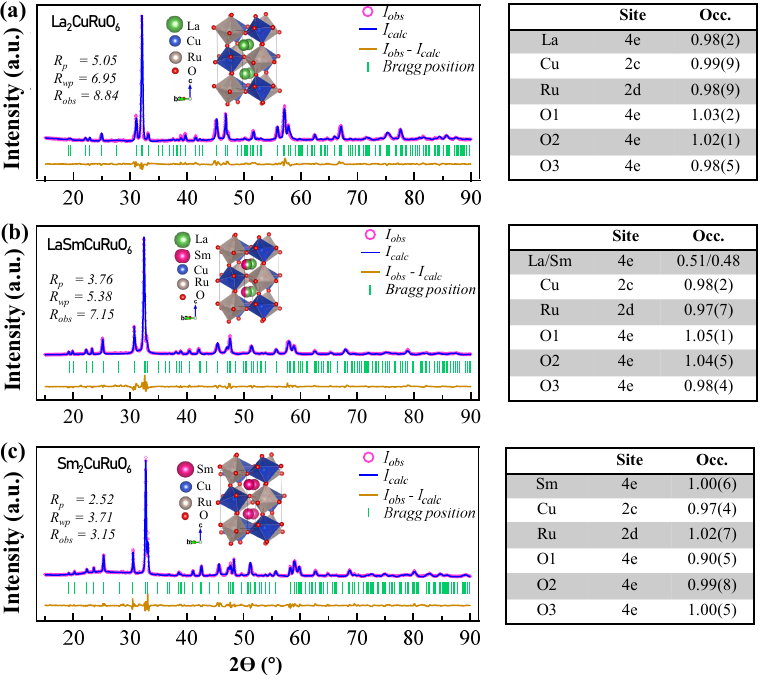}
    \caption{Rietveld refinement profile of the powder X-ray diffraction pattern for (a) La$_{2}$CuRuO$_{6}$ (b) LaSmCuRuO$_{6}$ (c) Sm$_{2}$CuRuO$_{6}$ by changing the occupancies of atoms.}
    \label{fig-OCC}
\end{figure*}

To find out possible deviations from stoichiometry we conducted XRD analysis by refining the atomic occupancies. Within experimental uncertainty, no significant deviation from nominal composition was detected, as shown in Fig.~\ref{fig-OCC}. Thus, the final models were constrained to ideal stoichiometry.

\textcolor{black}{\section{X-ray Photoelectron Spectroscopy}}
\begin{figure*}[]
	\includegraphics[width=0.8\linewidth, clip=true]{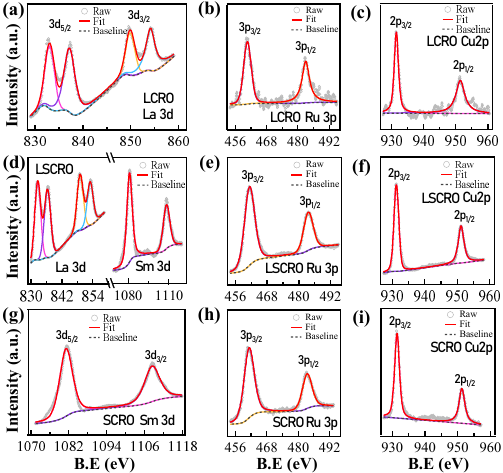}
	\caption{\textcolor{black}{(a) - (c) XPS spectra of La 3d, Ru 3p, and Cu 2p from La$_{2}$CuRuO$_{6}$ (LCRO). (d) - (f) XPS spectra of La 3d, Sm3d, Ru 3p, and Cu 2p from LaSmCuRuO$_{6}$ (LSCRO). (g) - (i) XPS spectra of Sm 3d, Ru 3p, and Cu2p from Sm$_{2}$CuRuO$_{6}$ (SCRO).}}
	\label{fig-XPS}
\end{figure*}
\textcolor{black}{X-ray photoelectron spectroscopy (XPS) was utilized to probe the oxidation states of transition-metal and rare-earth elements. The XPS spectra were analyzed using Shirley background correction to ensure accurate peak fitting. Figs.~\ref{fig-XPS}(a)-(c) present the XPS data for La 3d, Ru 3p, and Cu 2p states in LCRO, while Figs.~\ref{fig-XPS}(d)-(f) show the corresponding rare-earth and transition-metal element's XPS spectra for LSCRO, and Figs.~\ref{fig-XPS}(g)-(i) for SCRO. All the extracted binding energies, after fitting the corresponding XPS spectra, have been tabulated  in the Table~\ref{tab-XPS}. In the case of LCRO, the XPS data shows four prominent peaks for the La 3d states as shown in Fig.~\ref{fig-XPS}(a). The peaks at 833.08 eV and 837.22 eV are attributed to the La 3d$_{5/2}$, while the peaks at 849.86 eV and 854.10 eV correspond to La 3d$_{3/2}$. The splitting between these components arises due to a charge-transfer from the oxygen ligand to the unoccupied La 4f orbital~\cite{Mullica1985}. The binding energies observed for La are consistent with those reported earlier~\cite{Wang2021}, confirming that La maintains a $3+$ valence state in LCRO. Fig.~\ref{fig-XPS}(d) shows similar enegy positions of La 3d$_{5/2}$ and La 3d$_{3/2}$ for LSCRO with little variations, which are expected for slight changes in the atomic environment due to Sm substitution. This again confirms the $3+$ valence state of La in LSCRO. In Fig.~\ref{fig-XPS}(g), the XPS spectra for Sm 3d reveal two distinct peaks for SCRO. The Sm 3d$_{5/2}$ peak is observed at a binding energy of 1081.12 eV, while the Sm 3d$_{3/2}$ peak appears at 1108.43 eV, resulting in a spin-orbit splitting of 27.31 eV~\cite{Shit2024}. This energy separation is characteristic of the Sm$^{3+}$ oxidation state, confirming that samarium maintains a trivalent configuration in SCRO. Similarly, the energy position of Sm 3d$_{5/2}$ and Sm 3d$_{3/2}$ in LSCRO agrees well with the Sm$^{3+}$ oxidation state as can be seen in Fig.~\ref{fig-XPS}(d), and corresponding energy positions from Table~\ref{tab-XPS}.}

\begin{table*}[]
\centering
\caption{\label{tab-XPS} Binding energies extracted from XPS analysis. All the energies are in eV unit.}
\begin{tabularx}{\textwidth}{|>{\centering\arraybackslash}m{3cm}|>{\centering\arraybackslash}X|>{\centering\arraybackslash}X|>{\centering\arraybackslash}X|>{\centering\arraybackslash}X|>{\centering\arraybackslash}X|>{\centering\arraybackslash}X|>{\centering\arraybackslash}X|>{\centering\arraybackslash}X|}
\hline
\textbf{Material} & \textbf{La 3d$_{5/2}$} & \textbf{La 3d$_{3/2}$} & \textbf{Sm 3d$_{5/2}$} & \textbf{Sm 3d$_{3/2}$} & \textbf{Ru 3p$_{3/2}$} & \textbf{Ru 3p$_{1/2}$} & \textbf{Cu 2p$_{3/2}$} & \textbf{Cu 2p$_{1/2}$} \\
\hline
\textbf{La$_{2}$CuRuO$_{6}$} & 833.08,  837.22 & 849.86, 854.1 & - & - & 460.68 & 482.99 & 932.65 & 952.67 \\
\hline
\textbf{LaSmCuRuO$_{6}$} & 832.72, 836.60 & 849.51, 853.49 & 1080.95 & 1108.3 & 461.34 & 483.58 & 932.45 & 952.63 \\
\hline
\textbf{Sm$_{2}$CuRuO$_{6}$} & - & - & 1081.12 & 1108.43 & 461.22 & 483.43 & 933.22 & 953.46 \\
\hline
\end{tabularx}
\end{table*}

\textcolor{black}{The Ruthenium core level was analyzed by fitting the $Ru~3p$ doublet to provide a robust assessment of the Ru oxidation state. The fitted $Ru~3p_{3/2}$ and $Ru~3p_{1/2}$ binding energies are summarized in Table~\ref{tab-XPS} and displayed in Figs.~\ref{fig-XPS}(b), \ref{fig-XPS}(e), and \ref{fig-XPS}(h) for LCRO, LSCRO, and SCRO, respectively. The measured $Ru~3p_{3/2}$ positions across the series are consistent with the $Ru^{4+}$ oxidation state. The environment-driven structural changes due to substitution of La with Sm, alter the final-state screening and $Ru-O$ hybridization, leading to the small observed shifts in the core-level energy, which do not imply a change in the nominal $Ru^{4+}$ valence.}

\textcolor{black}{The Cu \(2p\) XPS spectra for all samples reveal two distinct peaks corresponding to the \(2p_{3/2}\) and \(2p_{1/2}\) spin-orbit splitting components, as can be seen from Figs.~\ref{fig-XPS}(c), \ref{fig-XPS}(f), and \ref{fig-XPS}(i). For LCRO, the \(2p_{3/2}\) peak appears at 932.65 eV, while the $2p_{1/2}$ peak is located at 952.67 eV. In SCRO, these peaks are slightly shifted to higher binding energies, with the \(2p_{3/2}\) at 933.22 eV and the $2p_{1/2}$ at 953.46 eV. The binding energies observed here are consistent with Cu in the $+2$ oxidation state, as expected for Cu$^{2+}$ in the octahedral coordination, commonly found in double perovskite oxides~\cite{Hu2012,Guo2018}. Additionally, a broad satellite feature observed around 942 eV, which is indicative of Cu$^{2+}$ $3d^9$ ground-state configuration~\cite{Fiermans1975,Qin2016}. The shifts in binding energies between LCRO and SCRO are attributed to the changes in the local electronic environment surrounding Cu, which occur due to the substitution of La with Sm at the A-site, influencing the Cu-O bond lengths within the CuO$_6$ octahedra and resulting in corresponding shifts in the Cu $2p$ binding energies~\cite{Hu2012,Guo2018}.}

Due to the significant spectral overlap of the Ru $3d$ doublet with the strong adventitious C $1s$ signal, we focused on the Ru $3p$ doublet for a robust assessment of the Ru core-level positions. The measured $Ru~3p_{3/2}$ and $3p_{1/2}$ binding energies are consistent across all three compositions (LCRO, LSCRO, and SCRO). These values align with the $Ru^{4+}$ oxidation state.

The Cu 2p spectra for all samples confirm the presence of the $Cu^{2+}$ oxidation state. A broad satellite feature is observed in the $940-943~eV$ range. This feature is a direct spectroscopic fingerprint for the $Cu^{2+}$ state, specifically indicating a $3d^9$ electronic configuration of $Cu^{2+}$ in the ground state~\cite{Fiermans1975}. It arises from an intrinsic $O~2p \rightarrow Cu~3d$ charge-transfer shake-up process during photoemission, and should not be interpreted as unreacted $\text{CuO}$ secondary phases. This is consistent with our main manuscript's X-ray diffraction (XRD) data, which shows a single-phase double perovskite structure. The improved signal-to-noise ratio in these measurements, subsequent to surface cleaning, presents this satellite as a broad, lower-intensity feature near the said energy region.

A systematic shift in the binding energy positions for both Cu $2p$ and Ru $3p$, as well as systematic variations in the spin-orbit splitting energies, are observed across the series as La is substituted by the smaller Sm. For Ru 3p, the spin-orbit splitting energy varies as 22.31 eV for LCRO, 22.24 eV for LSCRO, and 22.21 eV for SCRO. Similarly, for Cu 2p splitting energy varies as 20.02 eV for LCRO, 20.18 eV for LSCRO, and 20.24 eV for SCRO. These changes are not indicative of a change in formal valence. Instead, the substitution reduces the A-site radius, modifying the $Cu/Ru-O$ bond angles and octahedral tilting (as detailed in the main manuscript's XRD refinement). These environment-driven structural changes alter the nature of transition metal-oxygen bonding, affecting charge-transfer energies and final-state screening, thereby producing the observed subtle modifications in the core-level separation and position.

\section{Evolution of Coercivity \& Remanent Magnetization}

\begin{figure*}[h!]
    \centering
    \includegraphics[width=0.95\textwidth]{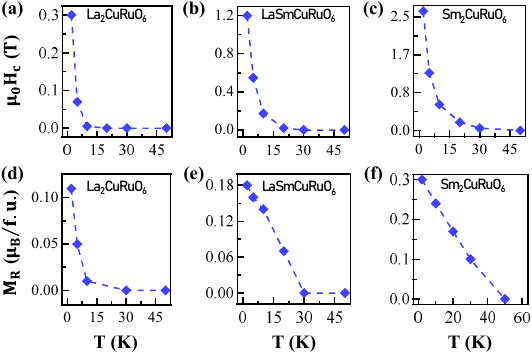}
    \caption{(a)-(c) Represent the temperature dependence of coercive field, whereas, (d)-(f) present the temperature dependence of remanent magnetization for La$_2$CuRuO$_6$, LaSmCuRuO$_6$, and (c)Sm$_2$CuRuO$_6$, respectively.}
    \label{fig-Hc-Mr}
\end{figure*}

The enhancement of coercivity and remanent magnetization as we go from La$_2$CuRuO$_6$ to Sm$_2$CuRuO$_6$, is shown in Fig.~\ref{fig-Hc-Mr}.

\textcolor{black}{\section{Heat Capacity}}

\begin{figure*}[]
\includegraphics[width=0.95\linewidth, clip=true]{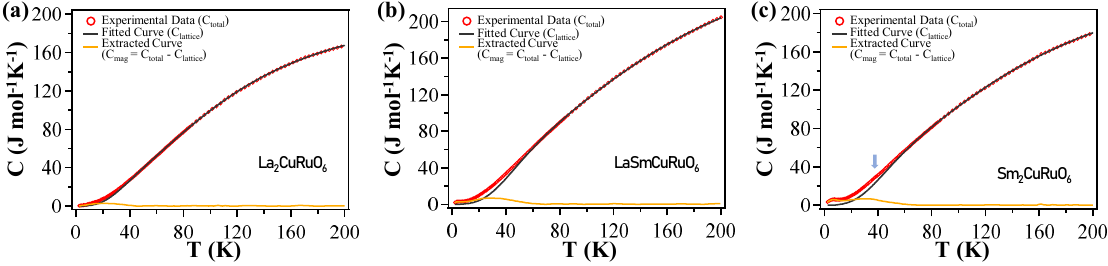}
	\caption{\textcolor{black}{Temperature dependent heat capacity (C) in the absence of magnetic field is shown for (a) $La_2CuRuO_6$, (b) $LaSmCuRuO_6$, and (c) $Sm_2CuRuO_6$.}}
	\label{fig-HC1}
\end{figure*}

\textcolor{black}{To understand the degree of magnetic ordering in these systems,  we measured the temperature dependence of the zero-field heat capacity, $C(T)$ as shown in Figs.~\ref{fig-HC1}(a), (b), and (c) for LCRO, LSCRO, and SCRO, respectively. These measurements allow for a detailed estimation of magnetic entropy and their possible magnetic ground state. To extract the magnetic contribution from the total heat capacity in SCRO, we attempted to model the lattice contribution using only the Debye term. However, this approach did not yield a satisfactory fit near the high-temperature region where magnetic contributions are negligible. We then consider one Debye term and one Einstein term, which reproduced the data very well [see Fig.~\ref{fig-HC1}(c)], using the equation,}

\begin{equation}
\textcolor{black}{C_{lattice}(T) = n_D\,C_D(T,\Theta_D) + n_E\,C_E(T,\Theta_E),}
\end{equation}
\textcolor{black}{here $n_D$ and ~$n_E$ are the fractional weights corresponding to the Debye and Einstein term, respectively. $C_D$ and $C_E$ are defined as,
}
\begin{equation}
\textcolor{black}{C_{D}(T, \Theta_D) \;=\; 9R \left(\frac{T}{\Theta_D}\right)^3
\int_{0}^{\Theta_D/T} \frac{x^4 e^x}{(e^x-1)^2}\,dx,}
\end{equation}

\begin{equation}
\textcolor{black}{and~ C_{E}(T, \Theta_E) \;=\; 3R \left(\frac{\Theta_E}{T}\right)^2
\frac{e^{\Theta_E/T}}{\left(e^{\Theta_E/T}-1\right)^2},}
\end{equation}
\textcolor{black}{where $R$ is the universal gas constant, $\Theta_D$ is the Debye temperature, and $\Theta_E$ is the Einstein temperature~\cite{Datta2023,Guchhait2025}. The best-fit parameters for SCRO are \(\Theta_E \approx 609\)~K, \(\Theta_D \approx 275\)~K and a combined fractional weights sum of $\sim$10, consistent with the number of atoms per formula unit.  After subtracting \(C_{lattice}\) from the measured $C(T)$, we obtain the magnetic specific heat \(C_{mag}(T)\) and the magnetic entropy was calculated by employing the equation,}
\begin{equation}
\textcolor{black}{S_{mag}(T)=\int \frac{C_{mag}(T)}{T}\,dT ,}
\end{equation}
\textcolor{black}{which saturates at $\sim$$15.9~J~mol^{-1} K^{-1}$ for SCRO. Following the same methodology, we also performed the heat capacity analysis of other two systems (LCRO and LSCRO), and the extracted numerical values from Debye-Einstein fitting alongside maximum magnetic entropy are tabulated in Tab.~\ref{tabHC}. The maximum magnetic entropy are $\sim$$12.9~J~mol^{-1} K^{-1}$ for LSCRO and $\sim$$4.2~J~mol^{-1} K^{-1}$ for LCRO.}

\begin{table}[]
\caption{\textcolor{black}{Numerical values extracted from Debye-Einstein fitting of heat capacity data and maximum magnetic entropy values.}}
\label{tabHC}
\begin{tabular}{lccc}
\toprule
Material & $\Theta_E$ (K) & $\Theta_D$ (K) & $S_{mag}$ (J\,mol$^{-1}$\,K$^{-1}$) \\
\midrule
La$_2$CuRuO$_6$ (LCRO)   & 456 & 235 & 4.2 \\
LaSmCuRuO$_6$ (LSCRO)    & 631 & 281 & 12.9 \\
Sm$_2$CuRuO$_6$ (SCRO)  & 609 & 275 & 15.9 \\
\bottomrule
\end{tabular}
\end{table}

\begin{figure*}[]
	\includegraphics[width=0.95\linewidth, clip=true]{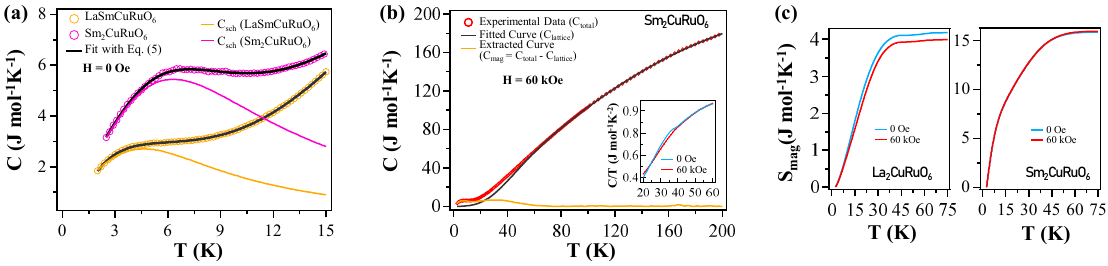}
	\caption{\textcolor{black}{(a) Low temperature zero-field heat capacity data for LaSmCuRuO$_{6}$, and Sm$_{2}$CuRuO$_{6}$ along with the fitted curve employing Eq. (5).} \textcolor{black}{(b) Temperature dependence of the heat capacity $C(T)$ for Sm$_{2}$CuRuO$_{6}$ under a 60~kOe applied field. Inset shows the plot of $C/T$ $vs.$ $T$ for Sm$_{2}$CuRuO$_{6}$ at 0 and 60~kOe, highlighting the suppression of the low-temperature anomaly by the applied  field. (c) Temperature-dependent magnetic entropy $S_{mag}$ calculated at 0 and 60~kOe for La$_{2}$CuRuO$_{6}$ (left panel) and Sm$_{2}$CuRuO$_{6}$ (right panel).}}
	\label{fig-HC2}
\end{figure*}

\textcolor{black}{The Boltzmann entropy for the transition-metal sublattice (Cu + Ru) can be calculated by the formula $Rln(2S+1)$. Considering localized Cu$^{2+}$ ($S=1/2$) and Ru$^{4+}$ ($S=1$) it will be $(R\ln2 + R\ln3 = R\ln6) \approx 14.90$~J\,mol\(^{-1}\)K\(^{-1}\). Now the release of zero field magnetic entropy, $S_{mag}$ for LCRO is only about 28\% of the total available entropy for this system. This is consistent with the formation of short-range correlations~\cite{Kumar2012,AnilKumar2017} and cluster-glass behaviour in LCRO and complement the findings of our AC susceptibility measurements. On the otherhand, SCRO displays a very different thermodynamic signature. Unlike non-magnetic La, Sm also may contribute to the total heat capacity. As a result, the extraction of the transition metal sublattice entropy is more complicated in SCRO.} \textcolor{black}{Since, there is a well-known crystal-field splitting of the Sm\(^{3+}\) \(J\)-multiplet and the associated Van Vleck paramagnetic contribution to the susceptibility,  only a small fraction of the full multiplet entropy is expected~\cite{Lopez2002,Harikrishnan2008,Mazumdar2021}. Therefore, to estimate the Sm contribution, we have analyzed the low temperature Schottky  anomaly region in the zeo-field heat capacity data of LSCRO, and SCRO [see Fig.~\ref{fig-HC2}(a)], using the equation,}

\begin{figure*}[]
	\includegraphics[width=0.7\linewidth, clip=true]{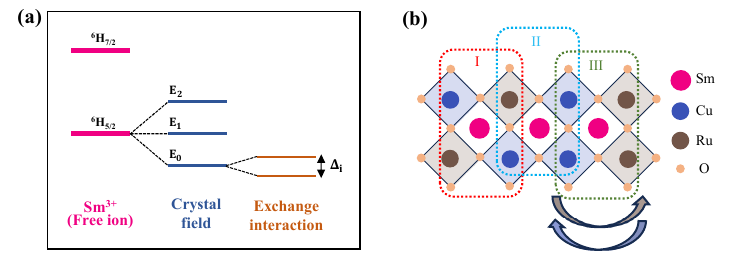}
	\caption{\textcolor{black}{(a) Schematic representation of $Sm^{3+}$ energy level splitting by low symmetry crystal-field effect and exchange interactions due to transition-metal sublattice environment. (b) Schematic representation of local Cu/Ru environment (I, II, or III) for Sm due to ASD.}}
	\label{fig-Sch}
\end{figure*}

\begin{equation}
\textcolor{black}{C(T) = C_{lat} + C_{mag} + C_{Sch}.}
\end{equation}
Here, the lattice heat capacity is given by $C_{lat} = \sum \beta_{2m+1} T^{2m+1}$ ($m=1,2,3,...$) at low temperatures~\cite{Lopez2002,Mazumdar2021}. To fit the data upto 15~K we have used ($m=2$), where $\beta_{3}$ parameter is related to the Debye temperature ($\theta_D$) by the equation $ \beta_{3} = \frac{12 \pi^{4} nR}{5 \, \Theta_{D}^{3}} $~\cite{Harikrishnan2008,Mazumdar2021}. The magnetic contribution can be expressed as $C_{mag} = \delta T^{p}$ ($p>0$), where $\delta$ is the stiffness-constant of spin-waves, and p equals to 1.5, and 3 for FM and AFM system respectively~\cite{Mazumdar2021,Patel2025}. Since, we are essentially working in the AFM region for these two systems, n is set to 3 for the low temperature fitting.
\textcolor{black}{The spin-orbit split ground state ($^6H_{5/2}$) of free $Sm^{3+}$ ion undergoes} a splitting into three Kramers doublet levels in a low symmetry crystal field \textcolor{black}{[cf. Fig.~\ref{fig-Sch}(a)], and at low temperatures only the lowest Kramers doublet ($E_0$) is mostly populated. The internal molecular field generated by the transition-metal sublattice environment [cf. Fig.~\ref{fig-Sch}(b)] lifts the degeneracy of this ground-state doublet ($E_0$) into two Zeeman levels [cf. Fig.~\ref{fig-Sch}(a)] giving rise to a Schottky anomaly. Furthermore, the presence of 10\% Cu/Ru ASD in the system suggests that the $Sm^{3+}$ ions do not reside in identical magnetic environment [cf. Fig.~\ref{fig-Sch}(b)]. To properly account this, the Schottky  anomaly is modeled as a superposition of three two-level systems representing the contribution of these varying local environments~\cite{Harikrishnan2008,Fertman2009,Mazumdar2021}:}

\begin{equation}
\textcolor{black}{ C_{Sch}(T) = N R \sum_{i=1}^{3}
   \left( {\omega_i}\frac{ \left( \frac{\Delta_i}{k_B T} \right)^2 \, e^{\Delta_i / (k_B T)} }
        { \left( 1 + e^{\Delta_i / (k_B T)} \right)^2 } \right).}
\end{equation}
Here, $N$ refers to the number of Sm$^{3+}$ ions per formula unit (i.e. N = 1 for LSCRO and N = 2 for SCRO). $\Delta_i$ is the \textcolor{black}{Zeeman} splitting energy of the ground state doublet \textcolor{black}{($E_0$), and the} weight factor $\omega_i$ ($\sum \omega_i = 1$) \textcolor{black}{represent the fractional probability of finding $Sm^{3+}$ ion in a particular local environment (I, II, or III) as shown in Fig.~\ref{fig-Sch}(b).}

So, the parameters extracted from fitting of low temperature data to the Eq. (5) have been tabulated in the Table~\ref{tabSch}. The extracted splitting energies are in good agreement with the Sm-based manganite perovskite system~\cite{Mazumdar2021}. If we calculate $\Theta_D$ from the $\beta_3$ parameter, the obtained value is 273.7 K for LSCRO, and 264.5 K for SCRO, which are in agreement to the high temperature Debye-Einstein fitting, i.e, 281 K for LSCRO, and 275 K for SCRO. We can also relate the splitting energy ($\Delta$) to the Schottky peak temperature ($T_S$)~\cite{Lopez2002} using the equation, $\Delta = \frac{k_B T_S}{0.418}$~\cite{Lopez2002,Mazumdar2021}. So, the energy splitting is 0.89 meV (LSCRO), and 1.36 meV (SCRO), which are in good agreement with the energy difference ($\Delta_3 - \Delta_1$) obtained from Schottky fitting, i.e., 0.92 meV (LSCRO), and 1.30 meV (SCRO). Therefore, after extracting the Schottky contribution ($C_{Sch}(T)$) for the above two samples we have calculated the corresponding value of entropy from the integration of $C_{Sch}/T$ to estimate the entropy contribution of $Sm^{3+}$. This method gives us $\sim 4.23~J\,mol^{-1}K^{-1}$ for LSCRO (one $Sm^{3+}/f.u.$), and $\sim 7.98~J\,mol^{-1}K^{-1}$ for SCRO (two $Sm^{3+}/f.u.$). Now, if we estimate the transition metal sublattice contribution to the magnetic entropy of SCRO, it will be \textcolor{black}{$\sim 7.9~J~mol^{-1} K^{-1}$}, which is about \textcolor{black}{53\%} of the total available entropy for transition metal sublattice, i.e., approximately two fold increase of entropy release in SCRO than in LCRO. This suggests a more cooperative magnetic interactions in SCRO than in LCRO.

\begin{table*}[]
\caption{\textcolor{black}{Extracted fitting parameters of the low temperature zero field heat capacity data using Eq. (5).}}
\label{tabSch}
\begin{tabular}{lccccccccc}
\toprule
Material & $\omega_1$ & $\omega_2$ & $\omega_3$ & $\Delta_1$ (meV) & $\Delta_2$ (meV) & $\Delta_3$ (meV) & $\Delta_3 - \Delta_1$ (meV) & $T_S$ (K) & $\Delta = \frac{k_B T_S}{0.418}$ (meV) \\
\midrule
LaSmCuRuO$_6$ (LSCRO)    & 0.16 & 0.35 & 0.50 & 0.31 & 0.53 & 1.23 & 0.92 & 4.3 & 0.89 \\
Sm$_2$CuRuO$_6$ (SCRO)  & 0.21 & 0.32 & 0.49 & 0.33 & 0.81 & 1.63 & 1.30 & 6.6 & 1.36 \\
\bottomrule
\end{tabular}
\end{table*}

We further explored the field-dependent heat capacity measurement on SCRO and LCRO. As shown in Fig.~\ref{fig-HC2}(b) for SCRO, the field-dependent $C(T)$ hardly show any change between zero field and with field in the high temperature region, except the kink at 37 K observed in zero field case is now suppressed when measured at 60 kOe [see inset in Fig.~\ref{fig-HC2}(b)].  This allow us to extract the magnetic heat capacity, $C_{mag}$ at 60 kOe by subtracting the zero-field Debye-Einstein lattice contribution from the total experimental $C(T)$ data. Further, using the integration method as detailed above, we subsequently derived the magnetic entropy ($S_{mag}$) for the 60 kOe data as shown in Fig.~\ref{fig-HC2}(c) for LCRO (left panel) and SCRO (right panel). We can clearly see that the applied external 60 kOe field reduced the overall magnetic entropy for LCRO, whereas it is almost unchanged in SCRO.

\section{Electrical Resistivity}

\begin{figure*}[ht]
\includegraphics[width=0.95\linewidth, clip=true]{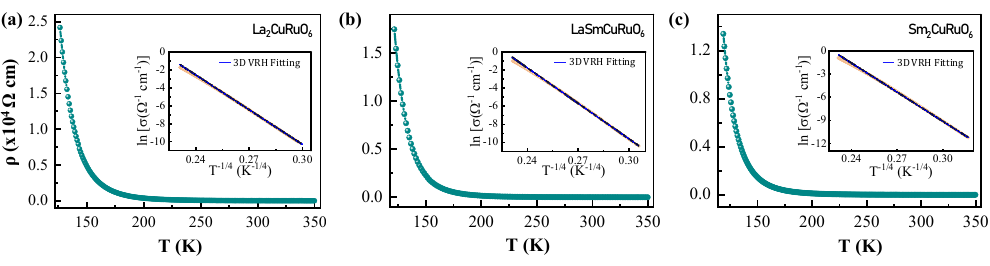}
	\caption{Temperature dependent resistivity for (a) La\(_2\)CuRuO\(_6\), (b) LaSmCuRuO\(_6\), and (c) Sm\(_2\)CuRuO\(_6\). The inset in each panel shows the fitting with variable range hopping beahviour using Eq. (1).}
	\label{fig-RT}
\end{figure*}

Figs.~\ref{fig-RT} (a)-(c) show the temperature-dependent electrical resistivity $\rho(T)$ of LCRO, LSCRO, and SCRO measured to investigate the electrical transport properties in these compounds. As shown in the insets, the plots of conductivity ($\sigma = 1/\rho$) as $\ln(\sigma)$ versus $T^{-1/4}$ are linear over the measured temperature range, confirming that the transport behavior in all samples follows Mott’s three-dimensional variable-range hopping (VRH) mechanism~\cite{Mott1969,Mott1979} over a broader temperature range. The VRH behaviour is typically observed in systems with localized charge carriers, and thermal excitation across an energy gap is insufficient for conduction. In such cases, charge carriers hop between localized states over varying distances, with the hopping probability dependent on temperature. For three-dimensional VRH, the conductivity is expressed by,

\begin{equation}
    \sigma(T) = \sigma_0 \exp\left[-\left(\frac{T_0}{T}\right)^{1/4}\right].
\end{equation}

Here $T_0$ is a characteristic temperature related to the density of localized states near the Fermi level. The fitted VRH parameters exhibit a systematic trend across the series: the characteristic temperatures $T_0^{1/4}$ are 129.45, 127.87, and 125.95 K$^{1/4}$ for LCRO, LSCRO, and SCRO, respectively. The effective activation energy in the VRH model is related to $T_0$ through the equation,

\begin{equation}
    E(T) = \frac{k_B}{4} T_0^{1/4} T^{3/4},
\end{equation}

and comparison of $E(T)$ at 300 K yields approximate activation energies of 201 meV, 198.5 meV, and 195.5 meV for LCRO, LSCRO, and SCRO, respectively.

Although all samples exhibit a high degree of B-site ordering within the monoclinic $P2_1/n$ symmetry, the presence of $\sim$10\% ASD arising from $Cu/Ru$ intermixing, provides the key microscopic origin of the observed VRH behaviour. In an ideally ordered LCRO double perovskite, Cu$^{2+}$ ($d^9$) and Ru$^{4+}$ ($d^4$) ions would occupy the alternate $B/B'$ sites perfectly, facilitating long-range superexchange and potentially more delocalized conduction. However, the $\sim$10\% ASD introduces randomly distributed $Cu–O–Cu$ and $Ru–O–Ru$ linkages, which locally break the superexchange network and act as random potential wells or barriers. These sites generate a distribution of localized electronic states near the Fermi level, suppressing coherent transport. As a result, the available thermal energy becomes insufficient for band conduction, and charge carriers instead hop between localized states over variable distances and energies, in accordance with Mott’s VRH mechanism~\cite{Mott1969,Mott1979}. Thus, the observed VRH behaviour in all three compounds arises as a direct consequence of the quantified ASD. The interplay between structural disorder and electronic localization provides a self-consistent explanation for the transport behavior observed across the three studied compounds.

\section{Magnetic moment of Sm and its role in the magnetic ground state of the SCRO compound}

In the main manuscript, the electronic and magnetic properties were investigated using the projector augmented wave (PAW) method within GGA+$U$ as implemented in VASP. In this framework, the Sm $4f$ states are treated as core states. In order to obtain a more reliable description of the localized $4f$ electrons and to explicitly evaluate the spin and orbital contributions to the Sm moment, we performed additional calculations using an all-electron full-potential linearized augmented plane-wave (FP LAPW) method including spin-orbit coupling (SOC).

The calculations have been performed within the framework of the GGA + $U$ approach together with the inclusion of SOC in a self-consistent manner. For Sm, the value of the Coulomb repulsion on-site has been varied from $U=4$ eV to $U=10$ eV, while a fixed Hund’s exchange interaction of $J=0.6$ eV was used. In case of transition metal ions, values of $U$ were considered as $9$ eV for Cu $3d$ and $3$ eV for Ru $4d$, whereas $J$ was kept fixed at $0.9$ eV for Cu and $0.8$ eV for Ru.

The obtained spin and orbital moments of Sm in the SCRO compound are shown in Table~\ref{tab:Sm_moments}. Irrespective of the value of $U$ considered and also more or less independent of the choice of the quantization axis, the spin moment $m_s$ stays almost fixed with the average value of around $m_s \sim 4.98~\mu_B$. On the other hand, the orbital moment exhibits a variation between $m_L \sim 2.3$ - $2.9~\mu_B$, which following expectation is oriented opposite to the spin moment. Thus, an effective net moment of the order of $2.1$ - $2.7~\mu_B$ emerges.


\begin{table*}[t]
\caption{Spin ($m_s$), orbital ($m_l$), and net magnetic moments of Sm in the SCRO compound for different values of Hubbard $U$ and spin quantization axes. Moments are given in $\mu_B$.}
\label{tab:Sm_moments}
\begin{ruledtabular}
\begin{tabular}{ccccc}
$U$ (eV) & Axis & $m_s$ ($\mu_B$) & $m_l$ ($\mu_B$) & Net moment ($\mu_B$) \\
\hline
4  & [001] & 4.99 & 2.27 & $\sim$2.7 \\
4  & [110] & 4.99 & 2.82 & $\sim$2.2 \\
7  & [001] & 4.98 & 2.35 & $\sim$2.6 \\
7  & [110] & 4.98 & 2.82 & $\sim$2.1 \\
10 & [001] & 4.98 & 2.53 & $\sim$2.4 \\
10 & [110] & 4.98 & 2.88 & $\sim$2.1 \\
\end{tabular}
\end{ruledtabular}
\end{table*}

\subsection{Exchange interaction parameters from FP LAPW calculations}

\begin{figure*}[]
\includegraphics[width=0.95\linewidth, clip=true]{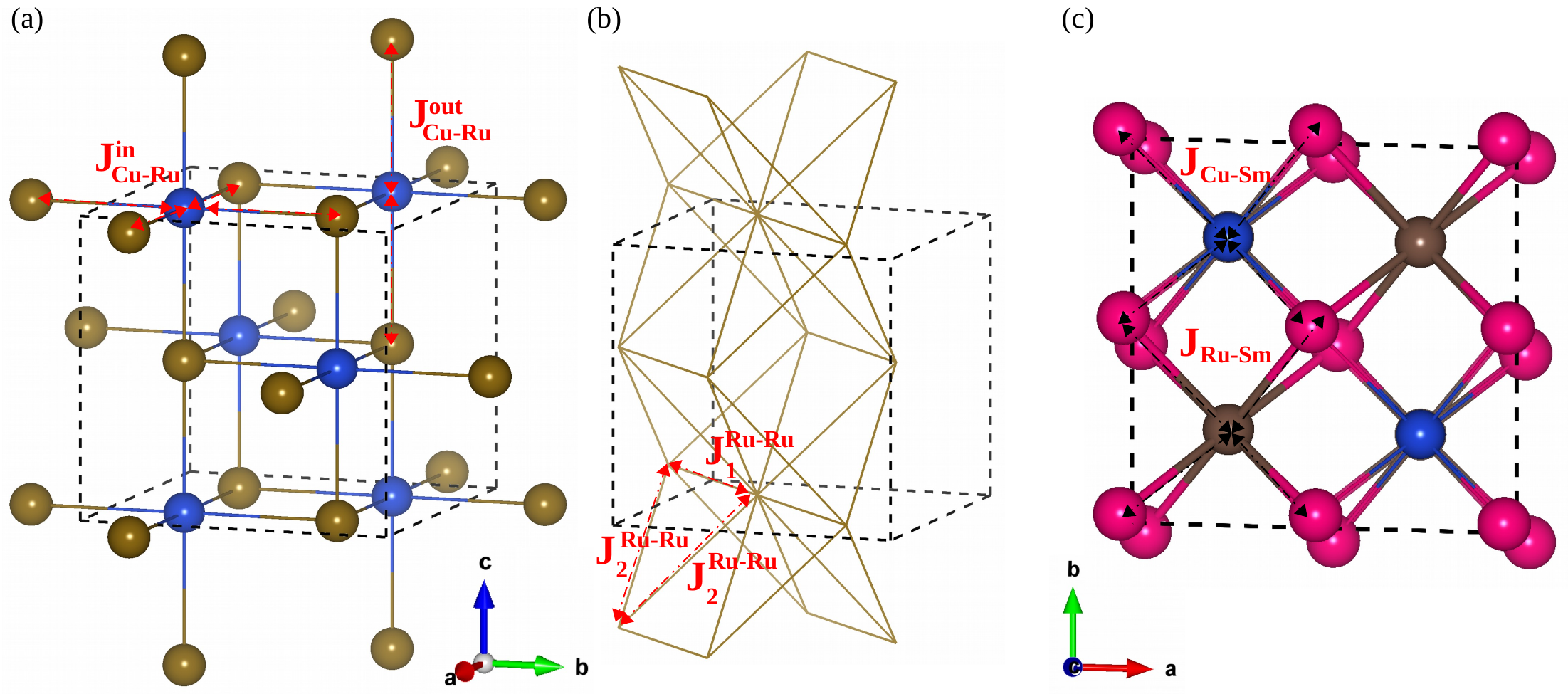}
\caption{Schematic illustration of the magnetic exchange pathways in the SCRO compound. (a) Nearest-neighbor Cu-Ru exchange interactions within the plane and along the out-of-plane direction, denoted as $J_{\mathrm{Cu-Ru}}^{\mathrm{in}}$ and $J_{\mathrm{Cu-Ru}}^{\mathrm{out}}$. (b) Next-nearest-neighbor Ru-Ru exchange paths, represented by $J_1^{Ru-Ru}$ and $J_2^{Ru-Ru}$, shown as a wireframe network where each node corresponds to a Ru ion. (c) Additional exchange pathways involving Sm, namely Cu-Sm and Ru-Sm interactions, illustrating the modification of the exchange network upon Sm substitution. Cu, Ru, and Sm ions are shown in blue, dark brown, and magenta, respectively. The inclusion of Sm introduces additional exchange channels and leads to a three-sublattice magnetic network in SCRO.}
\label{fig:exchange-path}
\end{figure*}

In order to shed light on the contribution of the Sm ions to the magnetic interactions, the magnetic exchange parameters have been obtained from total energy calculations done for different possible orientations of moments on Sm, Cu, and Ru within the FP LAPW scheme. A schematic illustration of the relevant exchange pathways in the SCRO compound is shown in Fig.~\ref{fig:exchange-path}. The calculated exchange parameters are summarized in Table~\ref{tab:exchange}. For the La-based compound (LCRO), only Cu-Ru and Ru-Ru interactions are present. In contrast, for the Sm-based compounds (SCRO and LSCRO), additional exchange paths involving Sm appear, namely Sm-Cu, Sm-Ru, and Sm-Sm. The differences in Cu-Ru and Ru-Ru interaction, presented in main text and
in Table~\ref{tab:exchange} arises due to difference in basis set, difference in atomic radii as well as implementation of GGA+U in the two codes (VASP and Wien2k).

\begin{table*}[]
\caption{Exchange interaction parameters (in meV) obtained from FP LAPW calculations including the Sm magnetic moment for SCRO, and LSCRO. For a consistent comparison on the same footing, calculations were also carried out for the La-based compound (LCRO). Negative (positive) values correspond to antiferromagnetic (ferromagnetic) interactions.}
\label{tab:exchange}
\begin{ruledtabular}
\resizebox{\textwidth}{!}{
\begin{tabular}{lccccccc}
Compound & $J_{\mathrm{Cu-Ru}}^{\mathrm{in}}$ & $J_{\mathrm{Cu-Ru}}^{\mathrm{out}}$ & $J_{1}^{\mathrm{Ru-Ru}}$ & $J_{2}^{\mathrm{Ru-Ru}}$ & $J_{\mathrm{Sm-Cu}}$ & $J_{\mathrm{Sm-Ru}}$ & $J_{\mathrm{Sm-Sm}}$ \\
\hline
SCRO      & -2.58 & -2.42 & -7.60 & -7.09 & -0.48 & 0.65 & 0.15 \\
LSCRO   & -1.93 & -1.87 & -7.48 & -7.32 & -0.20 & 0.30 & 0.05 \\
LCRO      & -1.68 & -1.65 & -7.44 & -7.41 & --    & --    & --   \\
\end{tabular}
}
\end{ruledtabular}
\end{table*}

A comparison of the interaction strengths shows that the dominant interaction in all cases is the Ru-Ru exchange, followed by the Cu-Ru interaction. \textcolor{black}{The exchange interactions involving Sm are about 10$\%$ of the dominant Ru-Ru interaction, while Sm-Sm are further smaller.
The presence of this additional magnetic degree
of freedom of Sm though modifies the overall magnetic network. Combined with the alignment of moments obtained from total energy
calculations (Sm parallel to Ru and antiparallel to Cu), this leads to a three-sublattice magnetic system in SCRO, in contrast to
the two-sublattice case in LCRO.
Inclusion of addition magnetic exchange contributions involving Sm in
a magnetically frustrated Sm-free system, becomes
effective in releasing the frustration.
This supports the conclusion drawn that the reduction in frustration arises from a combination of magnetic and structural effects, with the latter playing a leading role.}

\bibliography{reference-Sm2CuRuO6}

\end{document}